\documentclass[aps,pre,twocolumn,showpacs,showkey,square,numbers,amssymb,amsmath]{revtex4-1}
\usepackage{bm}
\usepackage{times,float}
\usepackage{graphicx}
\usepackage[usenames,dvipsnames,svgnames]{xcolor}
\usepackage{hyperref}
\hypersetup{colorlinks=true, linkcolor=NavyBlue, citecolor=PineGreen,urlcolor=cyan}

\newcommand{\bC     }{\mbox{\boldmath$C$}}

\begin{document}	
\title{Analytic solution of the two-star model with correlated degrees}
\author{Ma\'ira Bolfe}
\address{Physics Department, Federal University of Santa Maria, 97105-900 Santa Maria, Brazil. }
\author{Fernando L. Metz}
\email[]{fmetzfmetz@gmail.com}
\address{Physics Institute, Federal University of Rio Grande do Sul, 91501-970 Porto Alegre, Brazil}
\address{London Mathematical Laboratory, 8 Margravine Gardens, London W6 8RH, United Kingdom}
\author{Edgar Guzm\'an-Gonz\'alez}
\author{Isaac P\'erez Castillo}
\email[]{iperez@izt.uam.mx}
\address{Departamento de F\'isica, Universidad Aut\'onoma Metropolitana-Iztapalapa, San Rafael Atlixco 186, Ciudad de M\'exico 09340, Mexico}

\begin{abstract}
  Exponential random graphs are important to model the structure of real-world complex
  networks. Here we solve the two-star model with degree-degree correlations in the sparse regime. The model
  constraints the average correlation between the degrees of adjacent nodes (nearest neighbors)
  and between the degrees at the end-points of two-stars (next nearest neighbors).
  We compute exactly the network free energy and show that this model undergoes a first-order transition to a condensed phase. For non-negative degree correlations between
  next nearest neighbors, the degree distribution inside the condensed phase has a single peak at the largest degree, while
for negative degree correlations between next nearest neighbors the condensed phase is characterized by a bimodal degree distribution.
  We calculate the degree assortativities and show they are non-monotonic
  functions of the model parameters, with a discontinuous behavior at the first-order transition.
  The first-order critical line terminates at a second-order critical
  point, whose location in the phase diagram can be accurately determined.
  Our results can help to develop more detailed models of complex networks with correlated degrees.
\end{abstract}

\maketitle

\section{Introduction}

Random graphs constitute the main tool to model the complex behavior of large empirical networks
observed in social, technological, and biological systems \cite{Newman10,dorogBook}. In random graph models a
network is typically represented by nodes that interact through edges.
Random graph theory leads to important insights into the structure of networks as well as on the dynamical processes occurring on them, such as the spreading
of diseases \cite{VanMieghem2012,Goltsev2012}, the stability of ecosystems
to perturbations \cite{Grilli2016,Neri2020}, and the dynamics of sparsely connected neurons \cite{Brunel2000,Ostojic2014}.

A fruitful approach to network modeling consists of measuring a set of observables in an empirical network and then building an ensemble
of random graphs that matches these features in an average sense \cite{CoolenBook,Cimini2019}.
The probability distribution of graph configurations 
is derived by maximizing the network entropy subject to the constraints dictated
by the empirical observations \cite{Park2004b}. The resulting family of models, known as {\it exponential random graph} (ERG) models, aims to
reproduce a set of empirical patterns while keeping other network properties entirely
random. ERGs were introduced in the pioneering work of Holland and Leinhardt \cite{Holland1981}, and they soon
became popular models in social network analysis \cite{Frank1986,Strauss1986, Wasserman1996,Anderson1999,Snijders2011}. 

There are at least two main motivations to study ERGs. First, they serve as benchmark models to distinguish between random and
non-random traits in the structure of real-world networks \cite{Milo2002,Maslov2004,Squartini2011}.
Analytic solutions of ERGs give detailed information on the expected values of structural observables and
their fluctuations, which can be directly compared with data from real-world networks. Second, ERG models may exhibit degenerate
configurations and phase transitions, i.e, abrupt changes in the macroscopic properties of the graph ensemble.
Discontinuous phase transitions can be a serious limitation in the generation of ERGs, since they prevent that
certain configurations, with the desired structural features, are sampled.
Analytic solutions of ERGs predict
the existence and location of phase transitions in the parameter space. 

ERG models with specific constraints have been widely studied through numeric \cite{Palla2004,Jeong2007,Gorsky2020} and
analytic \cite{Berg2002,Park2004a,Park2004b,Burda2004a,Burda2004b,Annibale2009,Annibale2015,Coolen2016,Cimini2019,Aguirre2020,Aguirre2020a} techniques from statistical mechanics.
The two-star model is probably the simplest of ERGs that undergoes a phase transition \cite{Park2004a,Park2004b}. In this model, the graph ensemble
is constrained by the average number of edges and the average number of two-stars (a two-star is a pair of edges that share a
common node). The two-star model has been originally solved in the high-connectivity regime \cite{Park2004a,Park2004b}, using mean-field
techniques, and more recently in the sparse regime \cite{Annibale2015}, where each node is
adjacent to a finite number of others.

Here we take the theory of ERGs one step further
by solving the two-star model with degree-degree correlations \cite{Newman2002,Newman2003}, which
is an important structural feature of networks.
In general, the degrees
in real-world networks are not independent, but they can be positively or negatively correlated
with each other, as quantified by the Pearson correlation coefficient \cite{Newman2002}.
Nodes with similar degrees have positive degree correlations, whereas nodes with highly distinct degrees
have negative degree correlations. For instance, the degrees of adjacent or nearest neighbor nodes
in social networks are
positively correlated  \cite{Newman2002,Newman2003}, but these correlations become negative for
pairs of nodes connected through paths with more than one edge  \cite{Mayo2015,Fujiki2018}.

Earlier works have focused on nearest neighbor degree correlations \cite{Newman2002,Newman2003} and their impact on dynamical processes on networks, such as the spreading
of diseases \cite{Eguiluz2002,Boguna2003} and the synchronization of coupled oscillators \cite{Sorrentino2006,Nadal2015,Jalan2016}.
However, models that only take into
account local properties \cite{Orsini2015,Fujiki2017} do not reproduce certain global features of networks, such as their community structure or the
distribution of the shortest path length. In fact, recent works \cite{Rybski2010,Mayo2015,Arcagni2017,Fujiki2018,Mizutaka2020,Fujiki2020} have shown
that long-range degree correlations between nodes separated by more than one edge are
important for the organization of networks at a global level. For instance, results suggest
that the fractal structure of scale-free networks requires long-range degree correlations between hubs \cite{Fujiki2017}.
Long-range degree correlations have been also observed in the airport transportation network of the United States \cite{Mayo2015}, transcriptional regulatory
networks \cite{Mayo2015}, coauthorship networks \cite{Fujiki2018}, and the Twitter network  \cite{Fujiki2020}. Therefore, the solution of ERG models that incorporate
degree correlations in a systematic way represents a significant progress in network modeling.

In this work we solve the two-star model with degree-degree correlations between nearest neighbors and
between next nearest neighbors in the sparse regime. The free energy is exactly calculated
thanks to the introduction of an upper cutoff in the degree sequence.
We show  that the phase diagram of the model exhibits a first-order
critical line, surrounded by a metastable region, in which the graph sampling process may get stuck in a
local minimum of the free energy. The first-order transition separates a phase
characterized by an approximate Poisson degree distribution
from a condensed phase, where the degree distribution strongly depends on the degree correlations.
We quantify the degree
correlations through the degree assortativity corresponding to nearest neighbor nodes \cite{Newman2002,Newman2003} and to next nearest
neighbor nodes, located at the end-points of two-stars. When the degree assortativity of next nearest neighbors is non-negative, the degree distribution in the condensed
phase is peaked at the maximum degree; when the assortativity of next nearest neighbors is negative, the  condensed
phase is characterized by a bimodal degree distribution.
Both assortativities are non-monotonic functions
of the model parameters and exhibit a non-analytic behavior at the first-order transition.
The main theoretical findings are well corroborated by Monte Carlo simulations.

In the next section we introduce the generic framework of ERG models. In section \ref{observables} we define the main
structural properties of interest, including the two assortativity coefficients. Section \ref{solving} explains
how the model is analytically solved using conventional techniques of statistical mechanics, and how the structural
properties follow from the free energy. The  numerical results, obtained from the solutions of the saddle-point equations, are
discussed in section \ref{resulta}, while in the last section we present some final remarks. The appendix discusses
the symmetry properties of the order-parameter functions.

\section{Exponential random graph models} \label{sec2}

A graph configuration of a network with $N$ nodes
can be represented by a realization of the $N \times N$ adjacency
matrix $\bC$. The entries of $\bC$ fully encode the network topology, i.e., the matrix
element $C_{ij}$ tells whether there is an edge joining nodes $i$ and $j$.
We consider undirected
and simple random graphs \cite{Newman10}, which means that $\bC$ is a symmetric matrix with all diagonal elements equal to zero.
If there is an edge connecting nodes $i$ and $j$, then we set $C_{ij}=1$, whereas $C_{ij}=0$ otherwise. 
The degree $K_i$  of a node $i$
\begin{equation}
  K_i = \sum_{j=1}^N C_{ij}
  \label{popou}
\end{equation}
counts the number of edges attached to $i$, and the sequence $K_1,\dots,K_N$ contains
important information about the network structure. In this work we consider
random graph models in which the maximum
degree that may appear in a graph configuration is $k_{\rm max}$. The cutoff $k_{\rm max}$ is a model parameter, independent of $N$, which can be
freely adjusted.
As we will see below, the introduction of $k_{\rm max}$ allows to compute exactly the network properties
in the limit $N \rightarrow \infty$.

The probability $\mathcal{P}_N(\bC)$ to observe a certain graph configuration $\bC$
follows the Boltzmann-like form \cite{Park2004b}
\begin{equation}
  \mathcal{P}_N(\bC) =  \frac{ e^{-\mathcal{H}_N \left( \bC \right) }  }{\mathcal{Z}_N} \prod_{i=1}^N \Theta(k_{\rm max} - K_i)  ,
  \label{guraqo}
\end{equation}  
where $\Theta(x) =1$ if $x \geq 1$, and  $\Theta(x) =0$ otherwise.
The graph Hamiltonian $\mathcal{H}_N(\bC)$ depends on the network constraints, and $\mathcal{Z}_N$ is the
graph partition function
\begin{equation}
\mathcal{Z}_N = \sum_{\bC} e^{-\mathcal{H}_N \left( \bC \right)} \prod_{i=1}^N \Theta(k_{\rm max} - K_i)   .
\end{equation}  
The sum $\sum_{\bC}$  runs over all possible realizations of the adjacency matrix. To study the stability
of the graph configurations 
for $N \rightarrow \infty$, we need to compute the free energy density
\begin{equation}
  f = - \lim_{N \rightarrow \infty} \frac{1}{N} \ln \mathcal{Z}_N,
  \label{hupo}
\end{equation}  
which plays the role of a generating function for the  graph structural properties.

\section{Network observables} \label{observables}

The topology of graphs sampled from $\mathcal{P}_N(\bC)$ can
be characterized by a set of structural observables. In this work we only consider
global observables, which are obtained 
by averaging a local quantity over the entire network.

An important quantity to probe the network structure is the empirical degree distribution
\begin{equation}
  p_{k}(\bC) = \frac{1}{N} \sum_{i=1}^N \delta_{k,K_i},
  \label{pk}
\end{equation}  
which gives the probability that a randomly chosen node has degree $k$. The density of edges $\ell (\bC)$ and the density
of two-stars $s(\bC)$ are given by \cite{Park2004b,Annibale2015}
\begin{align}
\ell(\bC) &= \frac{1}{2 N} \sum_{i j=1}^N C_{ij} =  \frac{1}{2 N} \sum_{i=1}^N K_i, \label{por1} \\
s(\bC) &=  \frac{1}{2 N} \sum_{ijn=1}^N (1 - \delta_{in}) C_{ij} C_{jn} =     \frac{1}{2 N} \sum_{i=1}^N \left( K_i^2 - K_i \right).\label{por2} 
\end{align} 
A two-star (or path of length two) is a set with three different nodes $\{ i,j,k \}$ such that $C_{ij}C_{jk} = 1$.

Degree correlations are commonly quantified by the degree assortativity coefficient \cite{Newman2002,Newman2003,Mayo2015,Fujiki2018}. This is a global observable defined
as the Pearson correlation coefficient between the degrees of two nodes. Here we characterize the degree-degree correlations
by means of two assortativity parameters:
the standard assortativity $A^{(1)}(\boldsymbol{C})$ measures the degree correlations between adjacent nodes, while
the assortativity $A^{(2)}(\boldsymbol{C})$ measures the degree correlations between nodes at the end-points of two-stars.
In other words, $A^{(1)}$ ($A^{(2)}$)  quantifies degree correlations between nearest neighbors (next nearest neighbors).

In what follows, the indexes $k$ and $l$ refer to degrees. For a single graph instance, the assortativities are defined as
\begin{equation}
  A^{(r)}(\boldsymbol{C}) = \frac{\sum_{k l=0}^{\infty}  k l \, W_{k,l}^{(r)} (\boldsymbol{C})  - \left[  \sum_{k =0}^{\infty }  k  W_{k}^{(r)} (\boldsymbol{C})   \right]^2  }
  {  \sum_{k =0}^{\infty  }  k^2   W_{k}^{(r)} (\boldsymbol{C}) - \left[  \sum_{k =0}^{\infty }  k   W_{k}^{(r)} (\boldsymbol{C})  \right]^2   },
  \label{cuaop}
\end{equation}  
with $r = 1,2$.  The quantity
\begin{equation}
  W_{k,l}^{(1)} ( \bC ) = \frac{ \sum_{ij=1}^N C_{ij} \delta_{k,K_i} \delta_{l,K_j}   }{\sum_{ij=1}^N C_{ij} } 
  \label{gga11}
\end{equation}  
is the probability that
a randomly chosen edge joins two nodes with degrees $k$ and $l$, while 
\begin{equation}
  W_{k,l}^{(2)} ( \bC ) = \frac{ \sum_{ijn=1}^N \left( 1 - \delta_{in}  \right)  C_{ij} C_{jn}  \delta_{k,K_i} \delta_{l,K_n}   }{ \sum_{ijn=1}^N \left( 1 - \delta_{in}  \right)  C_{ij} C_{jn}    }
  \label{gga12}
\end{equation}  
is the probability that a randomly chosen two-star has degrees $k$ and $l$ at its end-points. The marginal
distributions
\begin{equation}
  W_{k}^{(r)} ( \bC ) = \sum_{l=0}^{\infty} W_{k,l}^{(r)} ( \bC ) \quad (r = 1,2)
\end{equation}  
have the explicit forms
\begin{align}
  W_{k}^{(1)} (\boldsymbol{C}) &=  \frac{1}{ 2 N \, \ell (\bC) } \sum_{i=1}^{N} K_i    \delta_{k,K_i} , \label{pouta1}  \\
  W_{k}^{(2)} (\boldsymbol{C}) &= \frac{1}{ 2 N \, s (\bC) } \sum_{i=1}^{N} \left(\sum_{j=1}^{N} C_{ij} K_j - K_i \right) \delta_{k,K_i} ,  \label{jupo}
\end{align}  
which can also be written as follows
\begin{align}
  W_{k}^{(1)} (\boldsymbol{C}) &= \frac{1}{2 \, \ell (\bC)} k \, p_{k}(\bC) , \\
  W_{k}^{(2)} (\boldsymbol{C}) &= \frac{\ell (\bC)}{s (\bC)} \sum_{l=0}^{\infty} (l-1) W_{k,l}^{(1)} ( \bC ).
\end{align}  
Equation (\ref{pouta1}) shows that the contribution of a node to $W_{k}^{(1)}$ is weighted according to
its degree, while the weight of node $i$ to the distribution $W_{k}^{(2)}$ is determined by the
number of edges attached to the neighbors of $i$, except from the links coming from $i$ itself.
This is intuitive, as a node $i$ with a large second-order degree $\sum_{j=1}^{N} C_{ij} K_j - K_i$ is
the end-point of a large number of two-stars.

The assortativities $A^{(1)}$ and $A^{(2)}$  give the same type of statistical information. Networks with statistically independent degrees satisfy
\begin{equation}
W_{k,l}^{(r)} ( \bC ) =  W_{k}^{(r)} (\boldsymbol{C})   W_{l}^{(r)} (\boldsymbol{C}),
\end{equation}  
and, consequently, $A^{(r)}(\boldsymbol{C}) = 0$. Networks with  $A^{(r)}(\boldsymbol{C}) > 0$ are positively correlated or assortative, which means that
nodes connected through edges or two-stars are likely to have similar degrees. Finally, networks with $A^{(r)}(\boldsymbol{C}) < 0$ are negatively correlated or disassortative, meaning that
nodes with large degrees preferentially connect through edges or two-stars to nodes with small degrees.

Equation (\ref{cuaop}) is not practical to calculate $A^{(r)}$. To prepare the ground for the analytic computation
of the assortativities, let us derive more convenient expressions for $A^{(1)}$ and $A^{(2)}$. Substituting Eqs.~(\ref{gga11}), (\ref{gga12}), (\ref{pouta1}), and (\ref{jupo}) in
Eq.~(\ref{cuaop}), we rewrite $A^{(1)}$ and $A^{(2)}$ as follows
\begin{align}
  A^{(1)}(\boldsymbol{C}) &= \frac{ \Lambda_{11}(\bC) - \left(\frac{1}{2 N \ell (\boldsymbol{C})}  \sum_{i=1}^N K_i^2  \right)^2}
  { \frac{1}{2 N \ell (\boldsymbol{C})}   \sum_{i=1}^N K_i^3 -  \left( \frac{1}{2 N \ell (\boldsymbol{C})}  \sum_{i=1}^N K_i^2  \right)^2      }, \label{tuio} \\
  A^{(2)}(\boldsymbol{C}) &=
  \frac{\chi(\boldsymbol{C}) - \left[  \Sigma \left( \boldsymbol{C} \right)  \right]^2}
       { \frac{\ell(\boldsymbol{C})}{s(\boldsymbol{C})} \Lambda_{21}(\bC) - \frac{1}{2 N s(\boldsymbol{C})  } \sum_{i=1}^N K_{i}^3
         - \left[ \Sigma \left( \boldsymbol{C} \right)   \right]^2  },
       \label{hudo}
\end{align}  
with 
\begin{equation}
\Sigma \left( \boldsymbol{C} \right) =   \frac{\ell(\boldsymbol{C})}{s(\boldsymbol{C})} \Lambda_{11}(\bC) - \frac{1}{2 N s(\boldsymbol{C})  } \sum_{i=1}^N K_{i}^2.
\end{equation}  
The object $\Lambda_{qr}(\bC)$ defines higher-order moments of nearest neighbor degrees
\begin{equation}
\Lambda_{qr}(\bC) = \frac{1}{ 2 N \ell (\bC) }  \sum_{ij=1}^N C_{ij} K_i^q K_j^r \quad (q,r \geq 1),
\end{equation}
and $\chi(\boldsymbol{C})$ is the  correlation between next nearest neighbor degrees
\begin{equation}
\chi(\boldsymbol{C}) = \frac{1}{2 N s \left(\boldsymbol{C}\right) } \sum_{ijn=1}^N \left( 1 - \delta_{in}  \right)  C_{ij} C_{jn} K_i K_n .
\end{equation}  
Equations (\ref{tuio}) and (\ref{hudo}) hold for a single realization of $\boldsymbol{C}$ and they show that
$A^{(1)}$ and $A^{(2)}$ are ultimately given in terms of moments of the joint distribution of degrees at different
pairs of nodes.

In the limit $N \rightarrow \infty$, the fluctuations
of intensive variables vanish and a single realization of an intensive quantity coincides
with its ensemble averaged value.
Thus, we naturally assume that the assortativities and all other global observables
of interest display such self-averaging behavior when $N \rightarrow \infty$, and the
assortativities become
\begin{align}
  \langle A^{(1)} \rangle &= \frac{ \left\langle \Lambda_{11} \right\rangle - \left(  \frac{\langle K^2  \rangle }{2 \langle \ell \rangle }   \right)^2}
  {  \frac{\langle K^3  \rangle }{2 \langle \ell \rangle }    -   \left(  \frac{\langle K^2  \rangle }{2 \langle \ell \rangle }   \right)^2     }, \label{tuio1} \\
  \langle A^{(2)} \rangle  &=
  \frac{\langle \chi \rangle - \left[ \frac{\langle \ell \rangle}{\langle s \rangle} \langle \Lambda_{11} \rangle  - \frac{\langle K^2 \rangle}{2 \langle s \rangle}   \right]^2}
       { \frac{\langle \ell \rangle }{\langle s \rangle}  \langle \Lambda_{21} \rangle - \frac{\langle K^3 \rangle }{2 \langle s \rangle   } 
         - \left[   \frac{\langle \ell \rangle}{\langle s \rangle} \langle \Lambda_{11} \rangle  - \frac{\langle K^2 \rangle}{2 \langle s \rangle}   \right]^2  },
       \label{hudo1}
\end{align}  
where $\langle \mathcal{G} \rangle$ denotes the ensemble average of an arbitrary random function  $\mathcal{G}(\bC)$ for $N \rightarrow \infty$
\begin{equation}
  \langle \mathcal{G} \rangle = \lim_{N \rightarrow \infty} \sum_{\bC} \mathcal{G}(\bC) \mathcal{P}_N(\bC).
  \label{upoita}
\end{equation}  
All ensemble averages in Eqs.~(\ref{tuio1}) and (\ref{hudo1}) can be calculated from the free energy $f$, which works
as a generating function for the moments of degrees.

\section{Analytic solution of the two-star model with correlated degrees} \label{solving}

In this section we present the Hamiltonian of the two-star model with correlated degrees and we explain how to
calculate,  in the limit $N \rightarrow \infty$, the free energy $f$ and the structural observables using standard tools from statistical mechanics.

\subsection{The Hamiltonian of the model}

The ERG model is defined by the Hamiltonian
\begin{align}
\mathcal{H}(\bC) = - \sum_{r=1}^Q \alpha_r \sum_{i=1}^N F_r(K_i)
- \frac{\gamma}{2} \sum_{i j=1}^N C_{ij}  D(K_i,K_j ) \nonumber \\
- \frac{\beta}{2} \sum_{ijk=1}^{N} \left(1 - \delta_{ik}   \right)  C_{ij} C_{jk} K_i K_k
+  \ln{N} \sum_{i < j} C_{ij},
\label{hamil}
\end{align}  
where $D(k,l)$ and
$F_1(k),\dots,F_Q(k)$ are arbitrary functions of the degrees, while $\beta$, $\gamma$, and $\alpha_1,\dots,\alpha_Q$ are
conjugate parameters that enforce the corresponding global constraints. The function $D(k,l)$ fulfills $D(k,l) = D(l,k)$.

From left to right, the first term in Eq.~(\ref{hamil}) enforces $Q$ global constraints involving single-site
functions $F_1(k),\dots,F_Q(k)$ of the degrees; the second term introduces a global constraint with $D(k,l)$ defined
at pairs of adjacent nodes; the third term couples the degrees of the next nearest neighbor nodes located
at the end-points of a two-star;
finally, due to the logarithmic scaling with $N$, the fourth term
in Eq.~(\ref{hamil}) ensures that networks sampled from $\mathcal{P}_N(\bC)$
are sparse \cite{Park2004b}, i.e., the probability of having an edge between two nodes
  is proportional to $1/N$ and the degrees $K_1,\dots,K_N$ remain
finite in the limit $N \rightarrow \infty$.

We obtain the two-star model with correlated degrees by setting
\begin{equation}
  F_r(k) = \delta_{r,1} k + \delta_{r,2} k^2, \quad D(k,l) = k l  .
\end{equation}
The motivation to solve the ERG model described by the Hamiltonian of Eq.~(\ref{hamil}) is twofold. First, the generic form of Eq.~(\ref{hamil}) allows to
calculate higher-order moments of the joint distribution of $K_1,\dots,K_N$ by  taking derivatives of  $f$ with respect to the conjugate parameters.
Such higher-order moments are needed to determine the assortativities $A^{(1)}$ and $A^{(2)}$ of the two-star model (see Eqs.~(\ref{tuio1}) and (\ref{hudo1})).
Second, although here we discuss explicit results for the two-star model
with correlated degrees, the flexible Hamiltonian of Eq.~(\ref{hamil}) allows to explore a variety of situations by combining
the simultaneous effect of different constraints.

\subsection{The calculation of the free energy}

In this subsection we solve the model defined by  Eq.~(\ref{hamil}). The aim is
to calculate the free energy $f$ in the limit $N \rightarrow \infty$, from which ensemble averages of
the network observables readily follow.

The graph partition function  reads
\begin{align}
  &\mathcal{Z}_N = \left(\prod_{i < j} \sum_{C_{ij} = 0,1}   \right)   \left[\prod_{i=1}^N \Theta\left( k_{\rm max} - K_i \right)    \right] e^{ -\ln N \sum_{i < j} C_{ij}   }  \nonumber \\
  &\times  \exp{\left( \sum_{r=1}^Q \alpha_r \sum_{i=1} F_r(K_i) + \gamma \sum_{i < j} C_{ij}  D(K_i,K_j) \right)} \nonumber \\
  &\times \exp{\left( \frac{\beta}{2} \sum_{ijk=1}^{N} \left(1 - \delta_{ik}   \right)  C_{ij} C_{jk} K_i K_k   \right)}.  \label{cucu2}
\end{align}
We remind that $K_1,\dots,K_N$ depend on the matrix elements $\{ C_{ij} \}$ according to Eq.~(\ref{popou}).
With the purpose of linearizing the exponent of Eq. (\ref{cucu2}) with respect to $\bC$, we rewrite
the above expression using Kronecker $\delta$'s
\begin{align}
  &\mathcal{Z}_N = \left(\prod_{i < j} \sum_{C_{ij} = 0,1}   \right)   \left[\prod_{i=1}^N \sum_{k_i=0}^{N-1} \delta_{k_i,K_i}  \Theta\left( k_{\rm max} - k_i \right)    \right] \nonumber \\
  &\times  \exp{\left( -\ln N \sum_{i < j} C_{ij}  \right)  }  \nonumber \\
  &\times  \exp{\left( \sum_{r=1}^Q \alpha_r \sum_{i=1} F_r(k_i) + \gamma \sum_{i < j} C_{ij}  D(k_i,k_j) \right)} \nonumber \\
  &\times \exp{\left( \frac{\beta}{2} \sum_{ijr=1}^{N} \left(1 - \delta_{ik}   \right)  C_{ij} C_{jr} k_i k_r   \right)}.  \nonumber
\end{align}
Using the integral representation
\begin{equation}
\delta_{k_i,K_i} = \int_{0}^{2 \pi} \frac{d u_i}{2 \pi} e^{i u_i (k_i - K_i)}
\end{equation}  
and substituting $K_i = \sum_{j=1}^N C_{ij}$, the partition function can be written as
\begin{align}
  &\mathcal{Z}_N = \left(\prod_{i < j} \sum_{C_{ij} = 0,1}   \right)   \sum_{k_1,\dots,k_N=0}^{k_{\rm max}}   \int_{0}^{2 \pi} \left( \prod_{i=1}^N \frac{du_i}{2 \pi} \right) \nonumber \\
  &\times  \exp{\left( i \sum_{i=1}^N k_i u_i  - i \sum_{i < j} C_{ij} (u_i + u_j) - \ln N \sum_{i < j} C_{ij}  \right)  }  \nonumber \\
  &\times  \exp{\left( \sum_{r=1}^Q \alpha_r \sum_{i=1} F_r(k_i) + \gamma \sum_{i < j} C_{ij}  D(k_i,k_j) \right)} \nonumber \\
  &\times  \exp{\left[ \frac{\beta}{2} \sum_{j=1}^{N} \left(\sum_{i=1}^N C_{ij} k_i    \right)^2  -  \frac{\beta}{2} \sum_{i j=1}^N C_{ij} k_i     \right]}.
  \label{gutoo}
\end{align}
It is still not possible to sum over the graph configurations, as the exponent in the
above equation contains a quadratic term in $\bC$.
We linearize this term via the exact identity
\begin{align}
 &\exp{\left[ \frac{\beta}{2} \sum_{j=1}^{N} \left(\sum_{i=1}^N C_{ij} k_i    \right)^2 \right]} = \int_{-\infty}^{\infty} 
  \left( \prod_{j=1}^N D x_j  \right) \nonumber \\
 &\times  \exp{\left(  \sqrt{\beta} \sum_{i < j} C_{ij}  \left( x_j k_i + x_i k_j   \right)  \right)  },   \label{poter}
\end{align}  
with the Gaussian measure
\begin{equation}
D x_j = \frac{d x_j}{\sqrt{2 \pi }} e^{  - \frac{1}{2}  x_j^2 }.
\end{equation}
Equation (\ref{poter}), known as the Hubbard – Stratonovich transformation, simply follows from a Gaussian integral \cite{NegeleBook}. 
Substituting Eq.~(\ref{poter}) in Eq.~(\ref{gutoo})
and summing over all graph configurations, we
arrive at an expression for $N \gg 1$
\begin{align}
 & \mathcal{Z}_N =  \sum_{k_1,\dots,k_N=0}^{k_{\rm max}} \int_{0}^{2 \pi} \left( \prod_{i=1}^N \frac{d u_i}{2 \pi}  \right) \int_{-\infty}^{\infty} \left( \prod_{j=1}^N D x_j  \right)
  \nonumber \\
  &\times
  \exp{\left( i \sum_{i=1}^N k_i u_i  + \sum_{r=1}^Q \alpha_r \sum_{i=1}^{N} F_r(k_i) \right)} \nonumber \\
&\times \exp{\left( \frac{1}{2 N}  \sum_{i j =1}^N e^{ - i (u_i + u_j) + \mathcal{W}_{\gamma,\beta}(k_i,x_i;k_j,x_j)  } \right)}, \label{poiuto} 
\end{align}  
where 
\begin{align}
  \mathcal{W}_{\gamma,\beta}(k,x;l,x^{\prime}) &=  \gamma D(k,l) -  \frac{\beta}{2} \left( k^2 + l^2 \right) \nonumber \\
  &+ \sqrt{\beta}   x^{\prime} k + \sqrt{\beta}  x l.
\end{align}

The last step is to decouple sites and reduce Eq.~(\ref{poiuto}) to a single-site
problem. This is achieved by introducing $k_{\rm max} + 1$ functional order-parameters
\begin{equation}
\rho_k(x) = \frac{1}{N} \sum_{i=1}^N \delta_{k,k_i} \delta(x-x_i) e^{- i u_i}, \quad  k=0,\dots,k_{\rm max},
\end{equation}  
through the following identity
\begin{align}
  &1 = \int \left( \prod_{k=0}^{k_{\rm max}} \mathcal{D} \rho_k \mathcal{D} \hat{\rho}_k \right)  \exp{\left( i \sum_{k=0}^{k_{\rm max}} \int d x \rho_k(x) \hat{\rho}_k(x) \right) } \nonumber \\
  &\times  \exp{\left( - \frac{i}{N}  \sum_{i=1}^N  \hat{\rho}_{k_i}(x_i)  e^{- i u_i} 
    \right) } ,
  \label{uopeq}
\end{align}  
where the functional integration measure is formally defined as
$\mathcal{D} \rho_k \mathcal{D} \hat{\rho}_k = \lim_{|\mathcal{X}| \rightarrow \infty} \prod_{x \in \mathcal{X} }  d \rho_k(x) d \hat{\rho}_k(x) / 2 \pi $, with
$\mathcal{X}$ representing the set of all possible values of $x$ obtained after discretization ($|\mathcal{X}|$ is the size of $\mathcal{X}$).
Inserting Eq.~(\ref{uopeq}) in Eq.~(\ref{poiuto}), we obtain
\begin{align}
 & \mathcal{Z}_N = \int \left( \prod_{k=0}^{k_{\rm max}} \mathcal{D} \rho_k \mathcal{D} \hat{\rho}_k \right)  \exp{\left( i \sum_{k=0}^{k_{\rm max}} \int d x \rho_k(x) \hat{\rho}_k(x) \right) } \nonumber \\
 &\times \exp{\left(  \frac{N}{2} \sum_{k, l =0}^{k_{\rm max}}  \int d x d x^{\prime}  \rho_k(x) \rho_{l}(x^{\prime})
    e^{ \mathcal{W}_{\gamma,\beta}(k,x;l,x^{\prime})  }     \right)} \nonumber \\
  &\times    \left(   \sum_{k=0}^{k_{\rm max}}   \int_{0}^{2 \pi}  \frac{d u}{2 \pi}  \int_{-\infty}^{\infty}  D x   e^{i k u + \sum_{r=1}^Q \alpha_r F_r(k)
          - \frac{i}{N}  \hat{\rho}_{k}(x)  e^{- i u} }  \right)^N.
\end{align}  
By rescaling the conjugate order-parameters as $\hat{\rho}_k(x) \rightarrow i N \hat{\rho}_k(x)$ and integrating over $u$, we find
a compact expression for $\mathcal{Z}_N$ when $N \gg 1$
\begin{equation}
   \mathcal{Z}_N =
    \int \left( \prod_{k=0}^{k_{\rm max}} \mathcal{D} \rho_k \mathcal{D} \hat{\rho}_k \right) \exp{\left(- N \mathcal{F}\left[\rho_{k},\hat{\rho}_{k} \right]    \right)}  ,
  \label{utar}
\end{equation}
in which 
\begin{align}
 &\mathcal{F}\left[\rho_{k},\hat{\rho}_{k} \right]  =  \sum_{k=0}^{k_{\rm max}} \int d x \rho_k(x) \hat{\rho}_k(x) \nonumber \\
 &- \frac{1}{2} \sum_{k,l =0}^{k_{\rm max}}  \int d x d x^{\prime}  \rho_k(x) \rho_{l}(x^{\prime})
  e^{ \mathcal{W}_{\gamma,\beta}(k,x;l,x^{\prime})  }     \nonumber \\
  &- \ln{   \left(   \sum_{k=0}^{k_{\rm max}}  \frac{1}{k!}  \int_{-\infty}^{\infty}  D x  
    \left[ \hat{\rho}_{k}(x) \right]^k   e^{ \sum_{r=1}^Q \alpha_r F_r(k) }    \right) }.
  \label{jojo}
\end{align}  
We have neglected the factor appearing in the integration measure of Eq. (\ref{utar}) due
to the rescaling $\hat{\rho}_k(x) \rightarrow i N \hat{\rho}_k(x)$, since this factor yields a subleading contribution
to the free-energy for large $N$. 
Equations (\ref{utar}) and (\ref{jojo}) determine the leading contribution to  $\ln \mathcal{Z}_N$ in
the limit $N \rightarrow \infty$. 

Since we introduced the finite cutoff $k_{\rm max}$ in the degree sequence, 
$\mathcal{F}\left[\rho_{k},\hat{\rho}_{k} \right]$ is independent of $N$ and, in the limit $N \rightarrow \infty$, the integral in Eq.~(\ref{utar}) can be solved through the
saddle-point method, according to which $\mathcal{Z}_N$ is dominated by the set of functions $\{ \rho_{k}^{*}(x), \hat{\rho}_{k}^{*}(x) \}$ that minimize
$\mathcal{F}\left[\rho_{k},\hat{\rho}_{k} \right]$. The fact the degrees are bounded in the present model is a crucial difference with respect to
reference \cite{Annibale2015}, which ensures the application of the saddle-point method and the convergence of the partition function for $N \rightarrow \infty$.
Thus, combining Eqs.~(\ref{utar}) and (\ref{hupo}), the free energy $f$ is directly given by
\begin{equation}
  f = \mathcal{F}\left[\rho_{k}^{*},\hat{\rho}_{k}^{*} \right],
  \label{free}
\end{equation}  
where $\{ \rho_{k}^{*}(x), \hat{\rho}_{k}^{*}(x) \}$ fulfills the saddle-point equations
\begin{align}
 & \hat{\rho}_{k}^{*}(x)  =  \sum_{l=0}^{k_{\rm max}}  \int d x^{\prime} \rho_{l}^{*}(x^{\prime}) e^{ \mathcal{W}_{\gamma,\beta}(k,x;l,x^{\prime})  }, \label{sad1} \\
  &  \rho_{k}^{*}(x) = \frac{   \frac{1}{ k!} k  \left[ \hat{\rho}_{k}^{*}(x) \right]^{k-1}   e^{ -\frac{1}{2} x^2 +  \sum_{r=1}^Q \alpha_r F_r(k) }      }{  \sum_{l=0}^{k_{\rm max}}  \frac{1}{l!}
    \int_{-\infty}^{\infty}  d x   \left[ \hat{\rho}_{l}^{*}(x) \right]^l   e^{ -\frac{1}{2} x^2 +   \sum_{r=1}^Q \alpha_r F_r(l) }    }, \label{sad2}
\end{align}  
with $k=0,\dots,k_{\rm max}$. The solutions of the self-consistent Eqs.~(\ref{sad1}) and (\ref{sad2}), together with the free
energy, Eq.~(\ref{free}), fully characterize the stability and the structural properties of infinitely large ERGs defined by Eq.~(\ref{hamil}).
We remark that Eqs.~(\ref{free}), (\ref{sad1}), and (\ref{sad2}) are exact in the limit $N \rightarrow \infty$.


\subsection{The equations for the structural observables}

In the limit $N \rightarrow \infty$, the ensemble averages of the network observables, defined in section \ref{observables}, follow from the derivatives of $f$ with respect to
the model parameters. Let us illustrate this fact by deriving the analytic expression for the degree distribution.
If we set $F_1(l) = \delta_{k,l}$ for arbitrary integers $0 \leq k, l \leq k_{\rm max}$, then the ensemble average
degree distribution $\langle p_k \rangle$ is determined from
\begin{equation}
  \langle p_k \rangle = - \frac{\partial f}{\partial \alpha_1},
\end{equation}  
which follows from Eqs. (\ref{guraqo}) and (\ref{hupo}).
From the explicit form of $f$, Eqs.~(\ref{jojo}) and (\ref{free}), we get
\begin{equation}
\langle p_k \rangle = \frac{  \frac{1}{k!}  \int_{-\infty}^{\infty}  D x  
    \left[ \hat{\rho}_{k}^{*}(x) \right]^k   e^{ \sum_{r=1}^Q \alpha_r F_r(k) }    }{    \sum_{l=0}^{k_{\rm max}}  \frac{1}{l!}  \int_{-\infty}^{\infty}  D x  
  \left[ \hat{\rho}_{l}^{*}(x) \right]^l   e^{ \sum_{r=1}^Q \alpha_r F_r(l) }  }.
\label{joqa}
\end{equation}  
This is a common strategy to calculate ensemble averages in statistical mechanics, namely, one performs the
derivative of the free energy with respect to a parameter that is coupled to a certain observable in the Hamiltonian.
Note that $F_1(k)$ in Eq.~(\ref{joqa}) is not necessarily given by $F_1(k) = \delta_{k,l}$. In other words, after the
choice $F_1(k) = \delta_{k,l}$ has served the purpose to obtain an expression for $\langle p_k \rangle$, we are free
to choose $F_1(k)$ as we please. 

Following an analogous procedure, the equations for the ensemble averages of all other observables
introduced in section \ref{observables} are obtained in the limit $N \rightarrow \infty$. The density of links and the density of two-stars read
\begin{align}
  &\langle \ell \rangle = \frac{\langle K \rangle}{2} , \label{re1} \\
  &\langle s \rangle = \frac{1}{2}  \left( \langle K^2 \rangle - \langle K \rangle \right), \label{re2}   
\end{align}  
where the moments $\langle K^{n} \rangle$ ($n=1,2,\dots$) of  $\langle p_k \rangle$
are determined from
\begin{equation}
\langle K^{n} \rangle = \sum_{k=0}^{k_{\rm max}} k^n \langle p_k \rangle.  
\end{equation}  
The moments $\langle \Lambda_{qr} \rangle$ of the joint degree distribution at adjacent nodes
read
\begin{align}
 &\langle \Lambda_{qr} \rangle = \frac{1}{2 \langle \ell \rangle} \sum_{kl=0}^{k_{\rm max}} k^q l^r   \int d x d x^{\prime}  \rho_{k}^{*}(x) \rho_{l}^{*}(x^{\prime})
  e^{ \mathcal{W}_{\gamma,\beta}(k,x;l,x^{\prime})  }, \label{zigui}
\end{align}
and the average degree correlation $\langle \chi \rangle$ at the end-points of two-stars is given by
\begin{align}
  \langle \chi \rangle &=  \frac{1}{2 \langle s \rangle} \sum_{kl=0}^{k_{\rm max}}  \int d x d x^{\prime}  \rho_{k}^{*}(x) \rho_{l}^{*}(x^{\prime}) \nonumber \\
  &\times
  \left( \frac{1}{\sqrt{\beta}} x l - k^2   \right)  e^{ \mathcal{W}_{\gamma,\beta}(k,x;l,x^{\prime})  }.
  \label{gonho}
\end{align}   
Once we determine $\{ \rho_{k}^{*}(x), \hat{\rho}_{k}^{*}(x) \}$ from the solutions of the
saddle-point Eqs.~(\ref{sad1}) and (\ref{sad2}), Eqs.~(\ref{joqa}-\ref{gonho}) allow
to compute the assortativities and characterize the network structure in the limit $N \rightarrow \infty$.
In case Eqs.~(\ref{sad1}) and (\ref{sad2}) have more than a single solution, the structural observables are calculated from
the solution that corresponds to the global minimum of $\mathcal{F}\left[\rho_{k},\hat{\rho}_{k} \right]$.


\section{Results} \label{resulta}

The exact equations derived in the previous section describe ERGs with the generic
Hamiltonian of Eq.~(\ref{hamil}) in the limit $N \rightarrow \infty$. In this section we
solve these equations and study the effect of degree correlations in the phase diagram of
the two-star model.

\subsection{The two-star model without degree correlations}

In this section we present results for the two-star model in the absence of degree correlations \cite{Park2004a,Park2004b,Annibale2015,Cimini2019,CoolenBook}, where the average density of edges
and the average density of two-stars are the only constraints. The model undergoes a discontinuous transition as a function of the control parameters both in the
dense regime \cite{Park2004a,Park2004b} and in the more realistic case of sparse networks \cite{Annibale2015}. In the latter case, reference \cite{Annibale2015} reports a
discontinuous behavior in the  structural parameters, but the stability of the macroscopic states and the corresponding  phase diagram remain elusive.
We complement the work of \cite{Annibale2015} by constructing
the full phase diagram of the two-star model in the sparse regime.

The Hamiltonian of the two-star model 
\begin{equation}
  \mathcal{H}(\bC) = - \alpha_1 \sum_{i=1}^N K_i  - \alpha_2 \sum_{i=1}^N K_i^2  +  \ln{N} \sum_{i < j} C_{ij}
  \label{gopqa}
\end{equation}  
is recovered from  Eq.~(\ref{hamil}) by setting
$\beta=\gamma=0$, $Q=2$, and $F_r(k) =\delta_{r,1} k + \delta_{r,2} k^2$. In this case, $\hat{\rho}_{k}^{*}(x)$ becomes
independent of $k$ and $x$, and we set $\hat{\rho}_{k}^{*}(x)  \equiv \mu$.
The fixed-point equation for $\mu$ follows from Eq.~(\ref{sad2})
\begin{equation}
\mu =  \frac{   \sum_{k=0}^{k_{\rm max}-1}  \frac{1}{ k!}  \mu^{k}   e^{ \alpha_1 (k+1) + \alpha_2 (k+1)^2    }      }{  \sum_{k=0}^{k_{\rm max}}  \frac{1}{k!}
  \mu^k   e^{ \alpha_1 k + \alpha_2 k^2 }    },
\label{pobuta}
\end{equation}  
and the free energy assumes the form
\begin{equation}
f =  \mathcal{F}(\nu) \big{|}_{\nu = \mu_{*}},
\label{tuiado}
\end{equation}  
where $\mu_{*}$  is the global
minimum of the function  $\mathcal{F}(\nu)$
\begin{equation}
\mathcal{F}(\nu) = \frac{1}{2} \nu^2 - \ln{\left(   \sum_{k=0}^{k_{\rm max}}  \frac{1}{k!}
  \nu^k   e^{ \alpha_1 k + \alpha_2 k^2 }  \right)}.
\end{equation}  
The degree distribution is obtained from Eq.~(\ref{joqa}) 
\begin{equation}
  \langle p_k \rangle = \frac{  \frac{1}{k!} \mu_{*}^k   e^{\alpha_1 k  +  \alpha_2 k^2 }    }
          {    \sum_{l=0}^{k_{\rm max}}  \frac{1}{l!}  \mu_{*}^l   e^{ \alpha_1 l  +  \alpha_2 l^2  }  }.
\label{joqa2}
\end{equation}  
In the limit $k_{\rm max} \rightarrow \infty$, Eqs.~(\ref{pobuta}) and (\ref{joqa2}) are
equivalent to the main equations in reference  \cite{Annibale2015}. However, in
contrast to \cite{Annibale2015}, the series in Eqs.~(\ref{pobuta}-\ref{joqa2})
contain a finite number of terms.
After $\mu_*$ is determined from the solutions of Eqs.~(\ref{pobuta}) and (\ref{tuiado}), $f$ and $\langle p_k \rangle$ can be evaluated for any $\alpha_1$ and $\alpha_2$.

\begin{figure}[h]
  \begin{center}
    \includegraphics[scale=0.6]{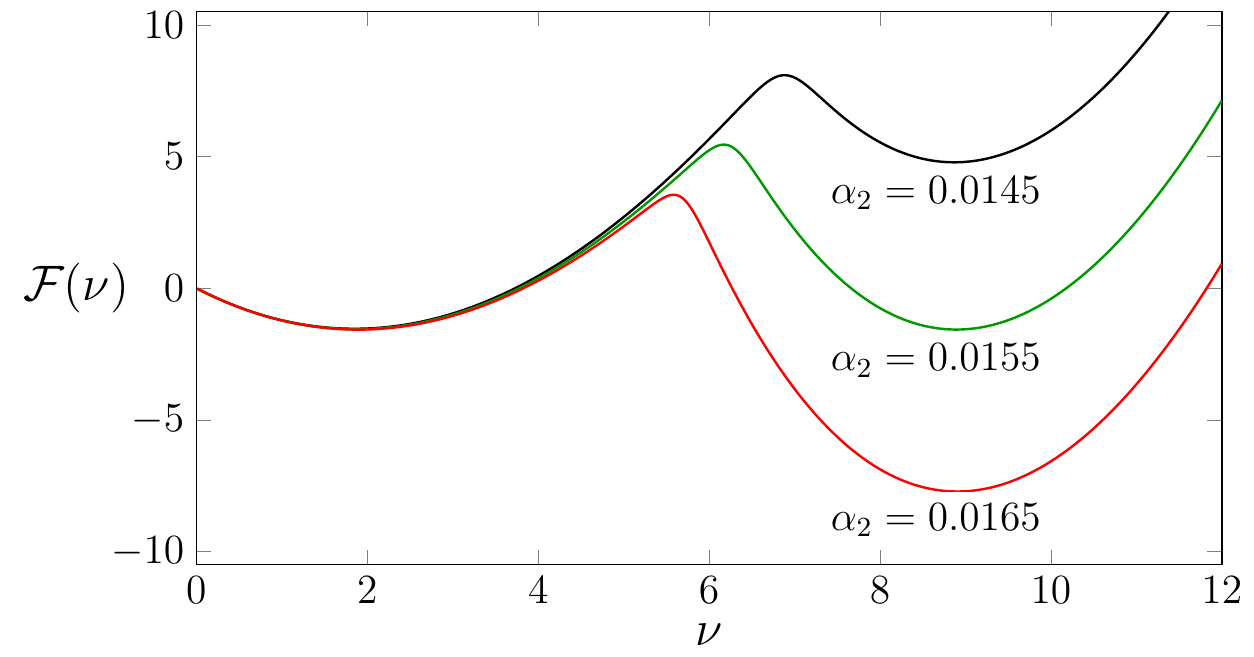}
    \caption{The function $\mathcal{F}(\nu)$ (see Eqs.~(\ref{utar}) and (\ref{jojo})) for the two-star model (see Eq.~(\ref{gopqa})) with
      maximum degree $k_{\rm max}=80$, $\alpha_1 = 0.5$, and different $\alpha_2$. The values of $\nu$ that minimize $\mathcal{F}(\nu)$
      are solutions of Eq.~(\ref{pobuta}). The global minimum determines the free energy and the structural parameters.
}
\label{fmu}
\end{center}
\end{figure}

Figure \ref{fmu} shows the functional behavior of $\mathcal{F}(\nu)$.
For the combinations of $\alpha_1$ and $\alpha_2$ shown in figure \ref{fmu}, $\mathcal{F}(\nu)$ exhibits two
minima, which reflects the existence of a metastable region in the phase diagram. Each minimum corresponds
to a stable fixed-point solution of Eq.~(\ref{pobuta}) and, consequently, to a certain macroscopic state of the two-star model.
The global minimum yields the leading contribution to the partition function for $N \rightarrow \infty$, from which one
determines the graph structural properties. The values of $(\alpha_1,\alpha_2)$ along which
the depths of the minima become equal identify a first-order critical line.
For fixed $\alpha_1$, $\mathcal{F}(\nu)$ has
a single minimum if $\alpha_2$ is either sufficiently large or small.
This particular situation is not shown
in figure \ref{fmu}.

\begin{figure}[h]
  \begin{center}
    \includegraphics[scale=0.95]{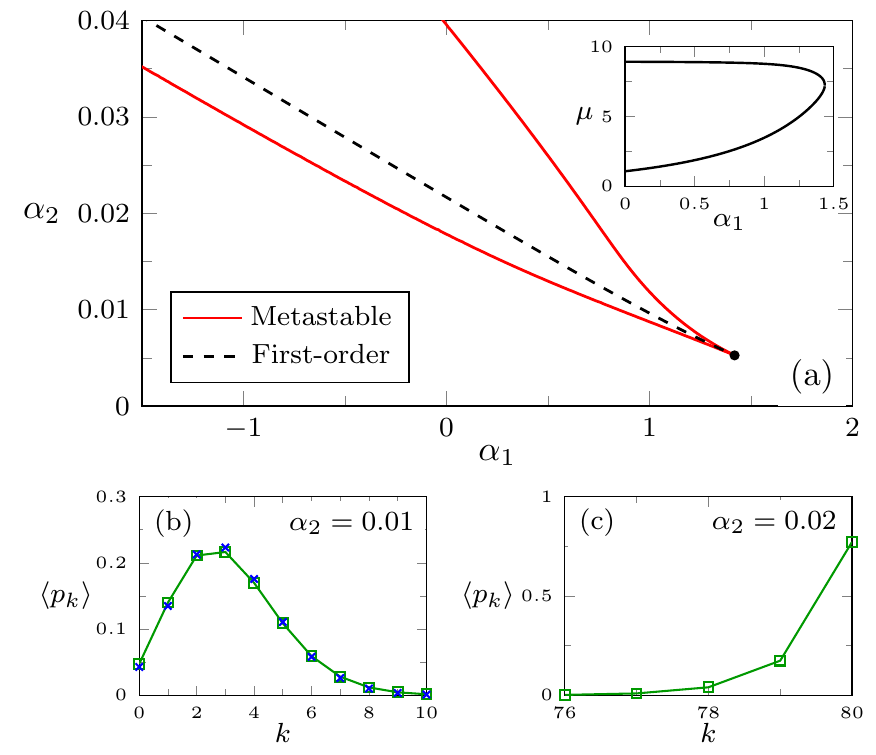}
    \caption{(a) Phase diagram of the two-star model with maximum degree $k_{\rm max}=80$ (see Eq.~(\ref{gopqa})). The dashed black
      curve is the first-order critical line and the solid red curves delimit the metastable region, within which the free energy has
      two minima. The inset shows the two stable solutions of Eq.~(\ref{pobuta}) as a function of $\alpha_1$ along
      the dashed curve. The two solutions for $\mu$ merge continuously at a critical point, identified by the black dot in figure (a).
      Figures (b) and (c) show the degree distribution (green squares)
      for fixed $\alpha_1=0.5$,  $k_{\rm max}=80$, and a value of $\alpha_2$ inside each phase. The blue crosses in figure (b)
      denote a Poisson distribution with the same mean degree.
}
\label{diag}
\end{center}
\end{figure}

The free energy allows to characterize the stability of the different phases and construct the phase diagram in the
plane $(\alpha_1,\alpha_2)$. The phase diagram for $k_{\rm max}=80$, shown in figure \ref{diag}-(a), exhibits a metastable
region enclosing a first-order critical line, which terminates at a critical point. The inset in figure  \ref{diag}-(a)
shows the  continuous phase transition of $\mu$ along the first-order critical line.
Figures \ref{diag}-(b) and \ref{diag}-(c) illustrate the typical profile of the degree distribution $\langle p_k \rangle$ in each phase.
Clearly, the first-order transition corresponds to an abrupt condensation of  $\langle p_k \rangle$ onto the maximum degree $k = k_{\rm max}$. Below the
first-order critical line, $\langle p_k \rangle$ is closer to a Poisson
distribution, whereas above the critical line $\langle p_k \rangle$ has a peak at $k = k_{\rm max}$
and the graph samples are approximately regular.

\begin{figure}[h]
  \begin{center}
    \includegraphics[scale=0.47]{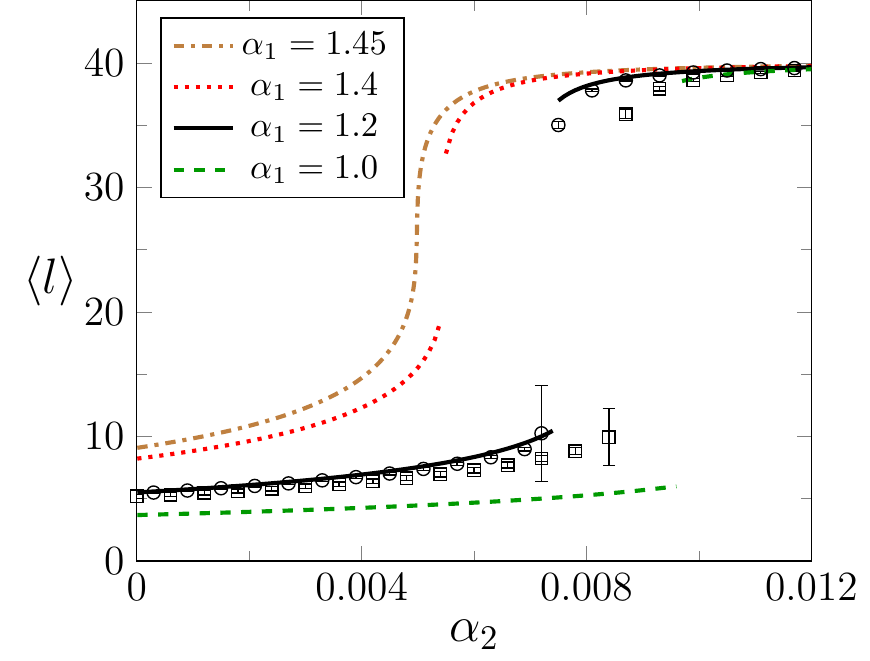}
    \includegraphics[scale=0.47]{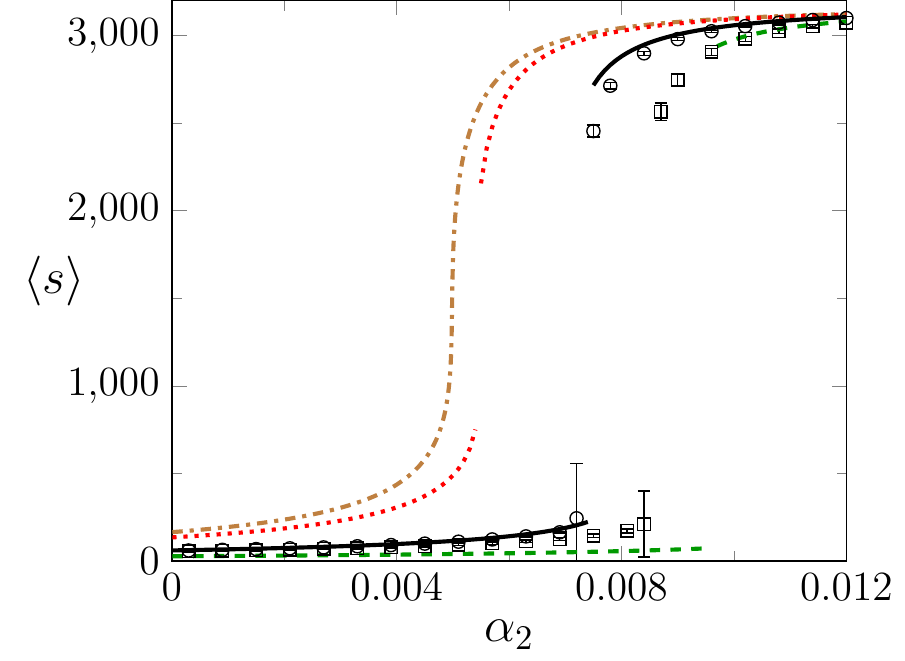}
    \caption{Theoretical results (different line styles) for the average density of edges $\langle \ell \rangle$  and the average density of two-stars
      $\langle s \rangle$ as a function of $\alpha_2$ for the two-star model  (see Eq.~(\ref{gopqa})) with maximum degree $k_{\rm max} = 80$
      and different $\alpha_1$. The symbols are obtained from the average over $10^5$ graph samples generated through Monte Carlo
      simulations with $\alpha_1=1.2$ and two system sizes: $N=200$ (squares) and $N=800$ (circles).       
    }
\label{lands}
\end{center}
\end{figure}

The behavior of the structural properties across the phase transition is a subject of practical interest.
Figure \ref{lands} shows the  density of links and the
density of two-stars as a function of $\alpha_2$ for different $\alpha_1$. Both
quantities are discontinuous at the first-order transition. The
discontinuity becomes gradually smaller as we increase $\alpha_1$, until it finally disappears at the
critical point, i.e., $\langle \ell \rangle$ and $\langle s \rangle$ are continuous and monotonic functions
of $\alpha_2$ provided $\alpha_1 \gtrsim 1.42$. Since $\beta=\gamma=0$, both assortativities are zero in this model. The theoretical results of figure \ref{lands} are
well corroborated by data obtained from Monte Carlo simulations that sample graphs from the distribution
of Eq. (\ref{guraqo}). Monte Carlo methods to generate ERGs are thoroughly discussed in \cite{CoolenBook}. 
 
\subsection{Degree correlations between nearest neighbors}

In this subsection we analyze the role of nearest neighbor degree correlations on the phase diagram of the two-star model. We consider an ERG model
that allows to tune the density of links, the density of two-stars, and the degree correlations between adjacent nodes.
The model is defined by the Hamiltonian
\begin{align}
  \mathcal{H}(\bC) &= - \alpha_1 \sum_{i=1}^N K_i - \alpha_2 \sum_{i=1}^N K_i^2    - \gamma \sum_{i < j}^N C_{ij} K_i K_j \nonumber \\
  &+  \ln{N} \sum_{i < j} C_{ij} ,
  \label{tuio2}
\end{align}  
which is recovered from  Eq.~(\ref{hamil}) by setting
$\beta=0$, $Q=2$, $F_r(k) =\delta_{r,1} k + \delta_{r,2} k^2$, and $D(k,l)=kl$. Most of the results in this section examine the effect of nearest neighbor degree correlations
when $\alpha_2=0$, since the results for $\alpha_2 \neq 0$ are qualitatively similar. Equation (\ref{tuio2}) for $\alpha_2=0$ is the Hamiltonian
of the Erd\"os-R\'enyi model \cite{Newman10} with degree correlations between adjacent nodes. We show below that degree correlations
induce a first-order condensation transition in the simple case of Erd\"os-R\'enyi random graphs.

The function $\hat{\rho}_k^{*}(x)$ is independent of $x$ for $\beta=0$. By writing
\begin{equation}
\int d x \rho_k^{*}(x) \equiv \rho_k^{*} , \quad \hat{\rho}_k^{*}(x) \equiv \hat{\rho}_k^{*},
\end{equation}  
the quantities $\{ \rho_k^{*} , \hat{\rho}_k^{*} \}_{k=0,\dots,k_{\rm max}}$ solve 
\begin{align}
 & \hat{\rho}_{k}^{*}  =  \sum_{l=0}^{k_{\rm max}}    \rho_{l}^{*} e^{\gamma k l   }, \label{sad12} \\
  &  \rho_{k}^{*} = \frac{   \frac{1}{ k!} k  \left( \hat{\rho}_{k}^{*} \right)^{k-1}   e^{ \alpha_1 k + \alpha_2 k^2 }      }
  {  \sum_{l=0}^{k_{\rm max}}  \frac{1}{l!} \left( \hat{\rho}_{l}^{*} \right)^l   e^{ \alpha_1 l + \alpha_2 l^2 }    }, \label{sad22}
\end{align}
while the free energy follows from
\begin{align}
 &f = \frac{1}{2} \sum_{k=0}^{k_{\rm max}}  \rho_k^{*} \hat{\rho}_k^{*}  - \ln{   \left(   \sum_{k=0}^{k_{\rm max}}  \frac{1}{k!} 
    \left( \hat{\rho}_{k}^{*} \right)^k   e^{ \alpha_1 k + \alpha_2 k^2}    \right) }.
  \label{jojorab}
\end{align}  
Equation (\ref{sad22}) represents a system of $k_{\rm max} + 1$ coupled fixed-point equations that can be solved
by iteration.

\begin{figure}[h]
  \begin{center}
    \includegraphics[scale=0.9]{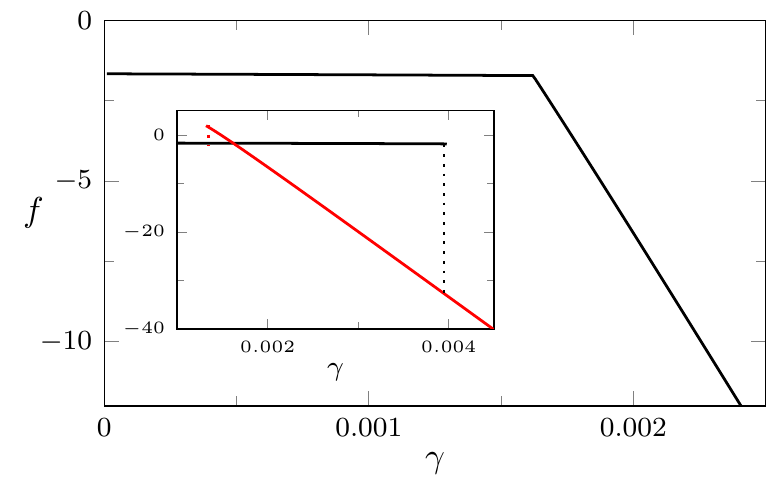}
    \caption{The main panel shows the free energy as a function of $\gamma$ for an exponential random graph model with nearest neighbor degree correlations (see Eq.~(\ref{tuio2})), maximum
      degree $k_{\rm max}=30$, $\alpha_1=0.6$, and $\alpha_2=0$. The non-analytic
      point of the free energy marks the first-order critical point. The inset displays the two branches of the function $f$ calculated from Eq. (\ref{jojorab}).
      The red solid line and the black solid line in the inset represent the function $f$ corresponding to each one of the two stable solutions of Eq.~(\ref{sad22}). The metastable
      region, where Eq.~(\ref{sad22}) has two stable solutions, is delimited by the dotted vertical lines. 
}
\label{fcorr1}
\end{center}
\end{figure}

The free energy is obtained from the global minimum of the $k_{\rm max} + 1$-dimensional
surface $\mathcal{F}\left( \rho_{0},\dots, \rho_{k_{\rm max}} \right) $ (see Eq.~(\ref{jojo})), which follows from the solutions of Eq. (\ref{sad22}).
Figure \ref{fcorr1} depicts $f$ as a function
of $\gamma$ for $\alpha_2=0$ and fixed $\alpha_1$. The free energy exhibits a non-analytic point, marking a first-order transition, at which the derivative
of $f$ with respect to $\gamma$ is discontinuous. The inset in figure \ref{fcorr1} shows the two branches of $f$, each one corresponding to
a minimum of $\mathcal{F}$ or a stable fixed-point of Eq. (\ref{sad22}).
In the metastable region, $\mathcal{F}\left( \rho_{0},\dots, \rho_{k_{\rm max}} \right)$ has two minima, one
of them is local (metastable), while
the other is global (stable). The structural properties of the
graph are evaluated from the solution $\{ \rho^{*}_k \}_{k=0}^{k_{\rm max}}$ at the global minimum of $\mathcal{F}$.
\begin{figure}[h]
  \begin{center}
    \includegraphics[scale=0.9]{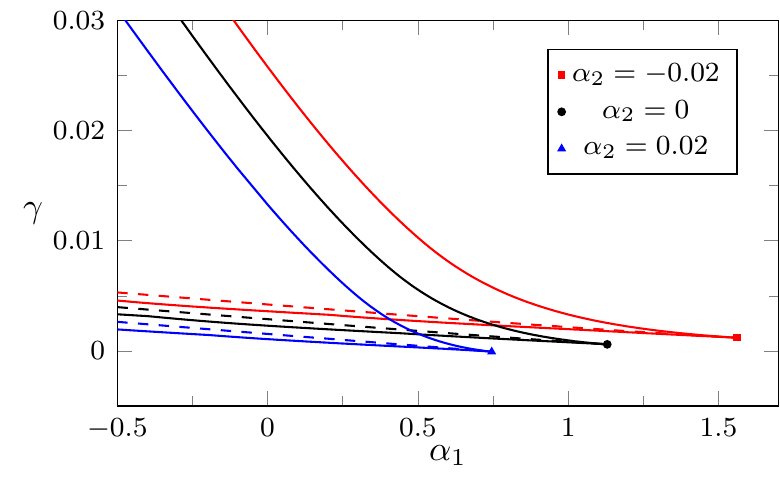}
    \caption{Phase diagram of the two-star model with nearest neighbor correlated degrees (see Eq.~(\ref{tuio2})), maximum degree $k_{\rm max}=30$, and different values
      of $\alpha_2$. The dashed curve for each $\alpha_2$ marks the first-order critical line and the solid curves delimit the metastable region, within which the free energy has
      two minima. For each value of $\alpha_2$, the dashed curve and the two solid curves terminate at the critical point identified by the corresponding symbol. The
      critical points are approximately given by $(\alpha_1,\gamma) = (1.56, 1.2 \times 10^{-3})$ for
      $\alpha_2=-0.02$, $(\alpha_1,\gamma) = (1.13, 6.1 \times 10^{-4})$ for $\alpha_2 = 0$, and $(\alpha_1,\gamma) = (0.746, -6 \times 10^{-5})$ for $\alpha_2 = 0.02$.
}
\label{diagcorr}
\end{center}
\end{figure}

The phase diagram $(\alpha_1,\gamma)$ of the ERG model defined by Eq.~(\ref{tuio2}), and obtained from the analysis of $f$, is shown in figure \ref{diagcorr}.
For each value of $\alpha_2$, the phase diagram has a first-order critical line, surrounded by a metastable region, which ends at a critical point. The first-order
critical line marks anew the condensation transition, above which the degree distribution $\langle p_k \rangle$ has a peak at $k = k_{\rm max}$.
The profile of $\langle p_k \rangle$ below and above each dashed line in figure  \ref{diagcorr} is qualitatively
similar to, respectively, figures \ref{diag}-(b) and \ref{diag}-(c). For $\gamma > 0$, adjacent nodes tend to have similar degrees, which strongly favors
the formation of a regular random graph, driving the ERG model to the condensed phase even in the case of $\alpha_2=0$.

\begin{figure}[h]
  \begin{center}
    \includegraphics[scale=0.53]{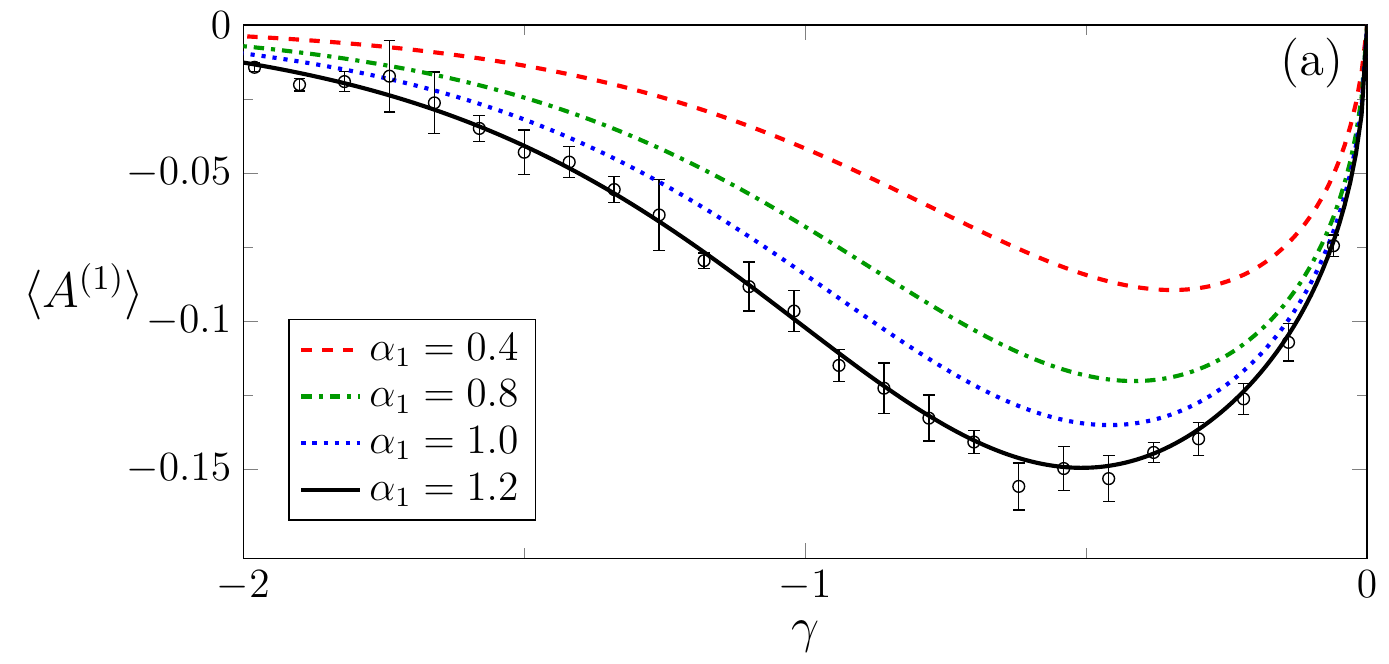} \\
    \hspace{0.1cm}
    \includegraphics[scale=0.53]{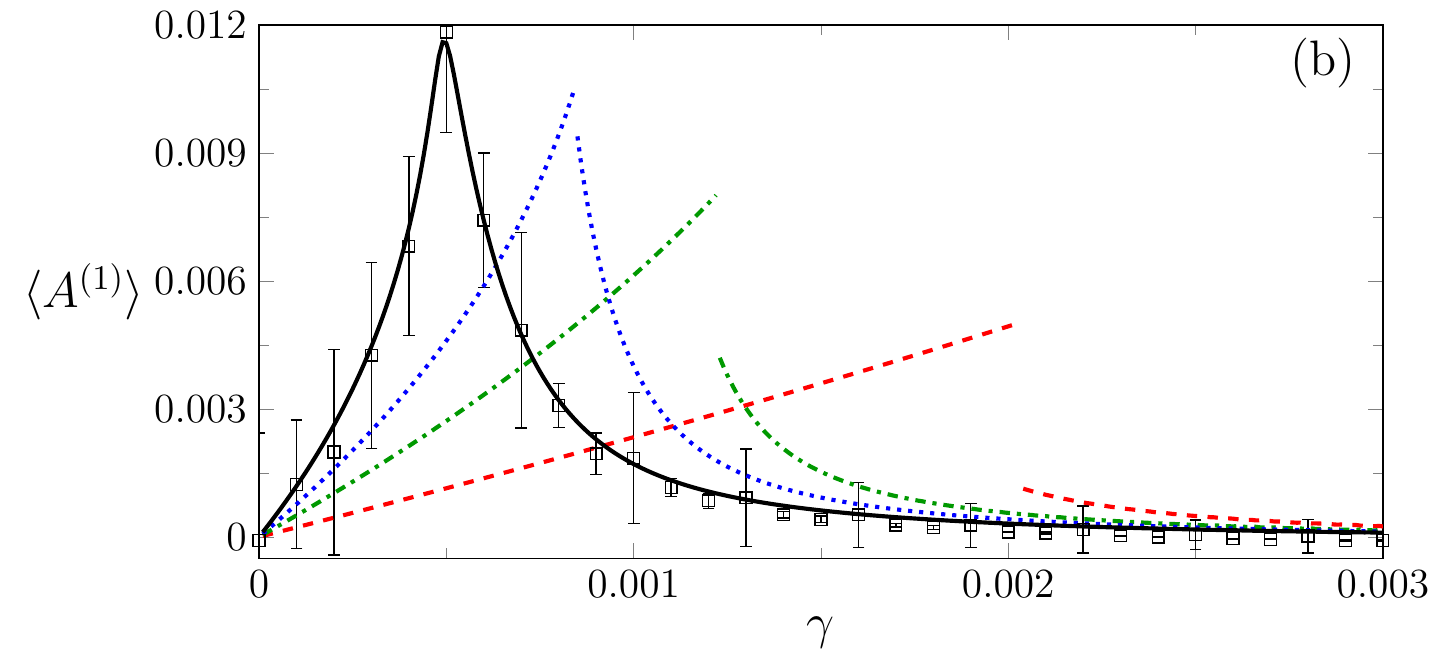}
    \caption{Theoretical results (different line styles) for the degree assortativity $\langle A^{(1)} \rangle$ of adjacent nodes as a
      function of $\gamma$ for an exponential random graph model with nearest neighbor correlated degrees  (see Eq.~(\ref{tuio2})), maximum degree $k_{\rm max} = 30$, and $\alpha_2=0$. The symbols
      are obtained from the average over $10^4$ graph samples generated through Monte Carlo simulations with $\alpha_2=0$, $\alpha_1 = 1.2$, $N=1500$ (circles), and $N=5000$ (squares). }
\label{assorta}
\end{center}
\end{figure}

The nearest neighbor assortativity $\langle A^{(1)} \rangle$ is discontinuous
at the first-order transition. Figure \ref{assorta} shows $\langle A^{(1)} \rangle$ as a function of $\gamma$ for different $\alpha_1$ and $\alpha_2=0$. The
discontinuity of $\langle A^{(1)} \rangle$ becomes smaller as $\alpha_1$ increases, until it vanishes continuously at
the critical point.
In fact, the two solutions for $\langle A^{(1)} \rangle$ corresponding to each phase merge into a single value as we
approach the critical point along the corresponding dashed curve in figure \ref{diagcorr}.
The assortativity $\langle A^{(1)} \rangle$ is a non-monotonic function of $\gamma$ that
vanishes when $|\gamma| \rightarrow \infty$. The latter property can be understood from the ground state configurations of the
Hamiltonian. For $\gamma \rightarrow -\infty$, the minimum of Eq.~(\ref{tuio2}) is attained when all degrees are zero; for
$\gamma \rightarrow \infty$, Eq.~(\ref{tuio2}) is minimized for all degrees equal to $k_{\rm max}$.
Since $\beta=0$, nodes at the end-points of two-stars are uncorrelated and $\langle A^{(2)} \rangle = 0$. Figure \ref{assorta} also shows results generated through  Monte Carlo
simulations of finite random graphs, which confirms our theoretical findings for $N \rightarrow \infty$.


\subsection{Degree correlations between next nearest neighbors}

In this subsection we focus on the competition between nearest neighbor degree correlations and next
nearest neighbor degree correlations. We consider the Hamiltonian
\begin{align}
\mathcal{H}(\bC) &=   \ln{N} \sum_{i < j} C_{ij} - \gamma \sum_{i < j}^N C_{ij} K_i K_j  \nonumber \\
&- \frac{\beta}{2} \sum_{ijk=1}^{N} \left(1 - \delta_{ik}   \right)  C_{ij} C_{jk} K_i K_k  ,
\label{hamil113}
\end{align}  
that results from Eq.~(\ref{hamil}) by setting $\alpha_r = 0$ for $r=1,\dots,Q$. We will not present explicit results for nonzero
values of $\alpha_1$ and $\alpha_2$ (see Eq. (\ref{gopqa})), since changing the density of links or two-stars does not modify the overall qualitative
picture discussed below.

In comparison to $\beta=0$, it is far more challenging
to solve the saddle-point Eqs. (\ref{sad1}) and (\ref{sad2})  and determine  the functions
$\rho_{0}^{*} (x),\dots,\rho_{k_{\rm max}}^{*} (x) $ for $\beta \neq 0$, due to the exponential or oscillatory behavior of the
integrands. We calculate numerically the integrals in Eqs. (\ref{sad1}) and (\ref{sad2}) by discretizing $\rho_{0}^{*} (x),\dots,\rho_{k_{\rm max}}^{*} (x)$
over $x \in \mathbb{R}$. The discretized version of the saddle-point equations is iterated until each function $\rho_{k}^{*} (x)$ converges
to a stationary functional form. In the appendix, we discuss some useful symmetry properties of $\rho_{0}^{*} (x),\dots,\rho_{k_{\rm max}}^{*} (x)$ for $\beta < 0$.
We set $k_{\rm max} =10$ throughout this subsection. 

\begin{figure}[h]
  \begin{center}
    \includegraphics[scale=0.53]{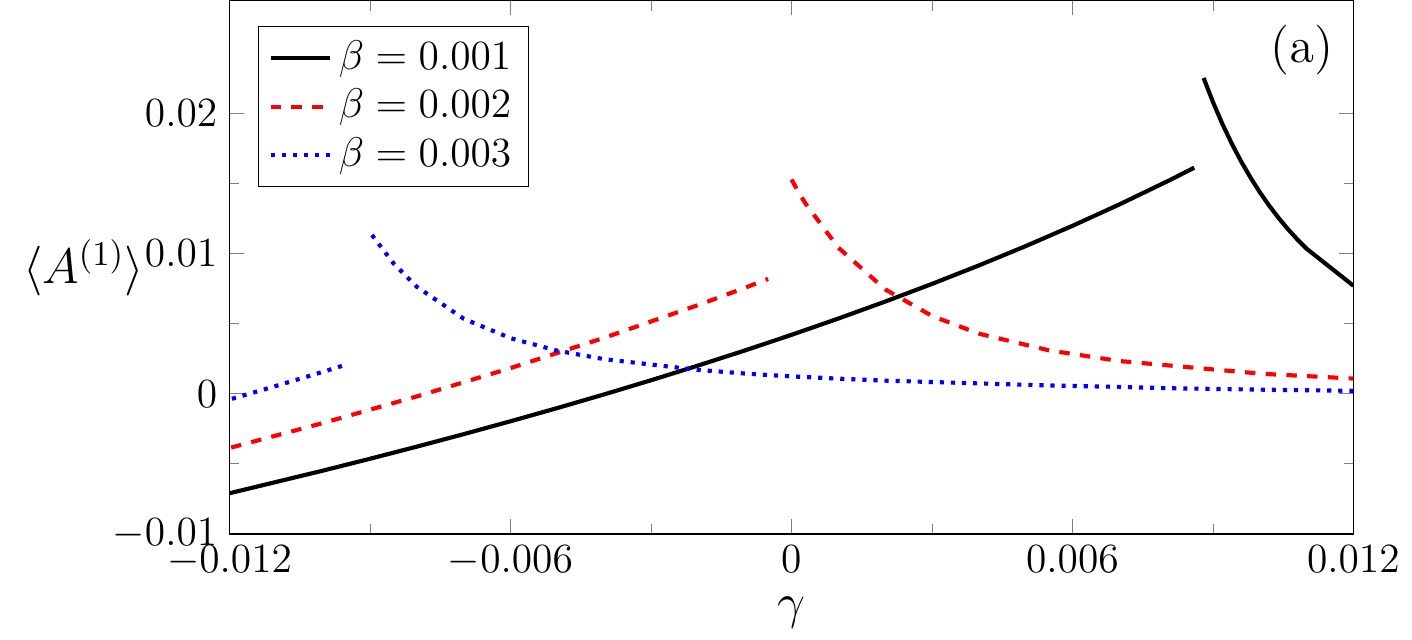} \\
    \hspace{-0.137cm}
    \includegraphics[scale=0.53]{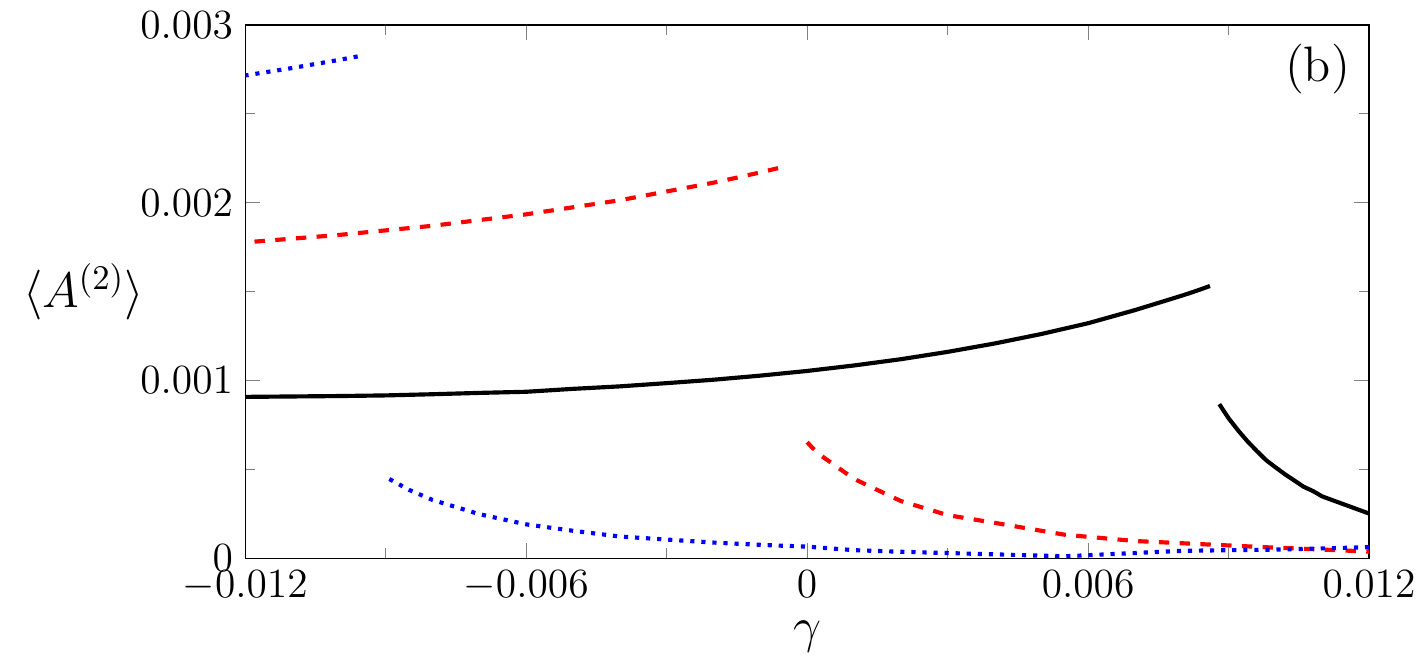}
    \caption{ Degree assortativities $\langle A^{(1)} \rangle$ and $\langle A^{(2)} \rangle$ corresponding, respectively, to nearest neighbor nodes and
      next nearest neighbors nodes of an exponential random graph model with degree-degree correlations (see Eq. (\ref{hamil113})), $\beta>0$, and maximum degree $k_{\rm max} = 10$.}
\label{assortbubu1}
\end{center}
\end{figure}
\begin{figure}[h]
  \begin{center}
    \includegraphics[scale=0.72]{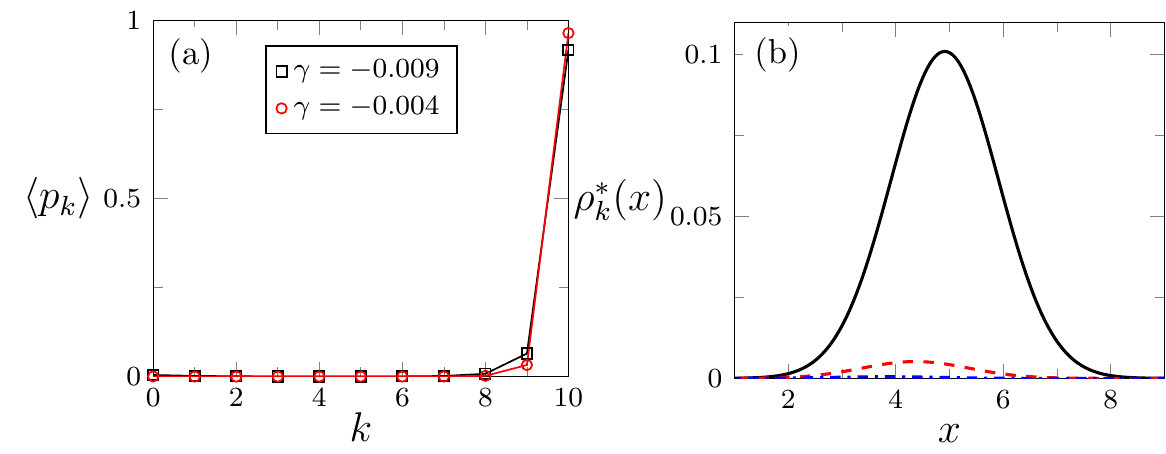}
    \caption{Degree distribution $\langle p_k \rangle$ and the order-parameter functions $\{ \rho_{k}^{*} (x) \}$ of an exponential random graph model with
      degree-degree correlations (see Eq. (\ref{hamil113})), $\beta = 0.003$, maximum
      degree $k_{\rm max} = 10$, and different values of $\gamma$ in the condensed phase.
      Figure (b) shows the functional behavior of $\rho_{10}^{*} (x)$ (black solid line), $\rho_{9}^{*} (x)$ (red dashed line), and
      $\rho_{8}^{*} (x)$ (blue dot-dashed line) for $\gamma = -0.004$. The other components $\rho_{0}^{*} (x), \dots, \rho_{7}^{*} (x)$
     are approximately zero.}
\label{assortatuta1}
\end{center}
\end{figure}

First, we consider the ERG model of Eq. (\ref{hamil113}) for $\beta > 0$. In this regime the order-parameters 
$\rho_{0}^{*} (x),\dots,\rho_{k_{\rm max}}^{*} (x)$ are real-valued functions.
Figure \ref{assortbubu1} shows that the degree 
assortativities $\langle A^{(1)} \rangle$ and $\langle A^{(2)} \rangle$
have a discontinuous behavior as a function of $\gamma$, and the model undergoes once more
a first-order transition at $\gamma=\gamma_c(\beta)$. For $\gamma < \gamma_c(\beta)$, $\langle p_k \rangle $ is approximately given by a Poisson
distribution, while for $\gamma > \gamma_c(\beta)$ the degree distribution 
has a single peak at $k=k_{\rm max}$. Figure \ref{assortatuta1} illustrates the behavior
of $\langle p_k \rangle $ and  $\rho_{0}^{*} (x),\dots,\rho_{k_{\rm max}}^{*} (x)$ inside the condensed phase.
For $\gamma \rightarrow \infty$ or $\beta \rightarrow \infty$, the degree distribution 
converges to $\langle p_k \rangle = \delta_{k,k_{\rm max}}$ and both assortativities are zero.
Overall, figures \ref{assortbubu1} and \ref{assortatuta1} show that positive next nearest neighbor degree correlations do not change
qualitatively the phase diagram.

Let us now present results for $\beta <0$, where $\rho_{0}^{*} (x),\dots,\rho_{k_{\rm max}}^{*} (x)$
are complex-valued functions of $x \in \mathbb{R}$. In the appendix, we demonstrate that ${\rm Re} \rho_{k}^{*} (x)$ is an even
function and ${\rm Im} \rho_{k}^{*} (x)$ is an odd function in the regime $\beta < 0$. It is interesting
to note that, for $\beta < 0$, we impose conflicting constraints in the generation of graph samples, since
negative values of $\beta$ favor dissimilar degrees at the end-points of two-stars, leaving the degrees at the central
nodes of two-stars in a frustrating situation. The appearance of these  frustrated configurations should influence the graph structure.

\begin{figure}[h]
  \begin{center}
    \includegraphics[scale=0.53]{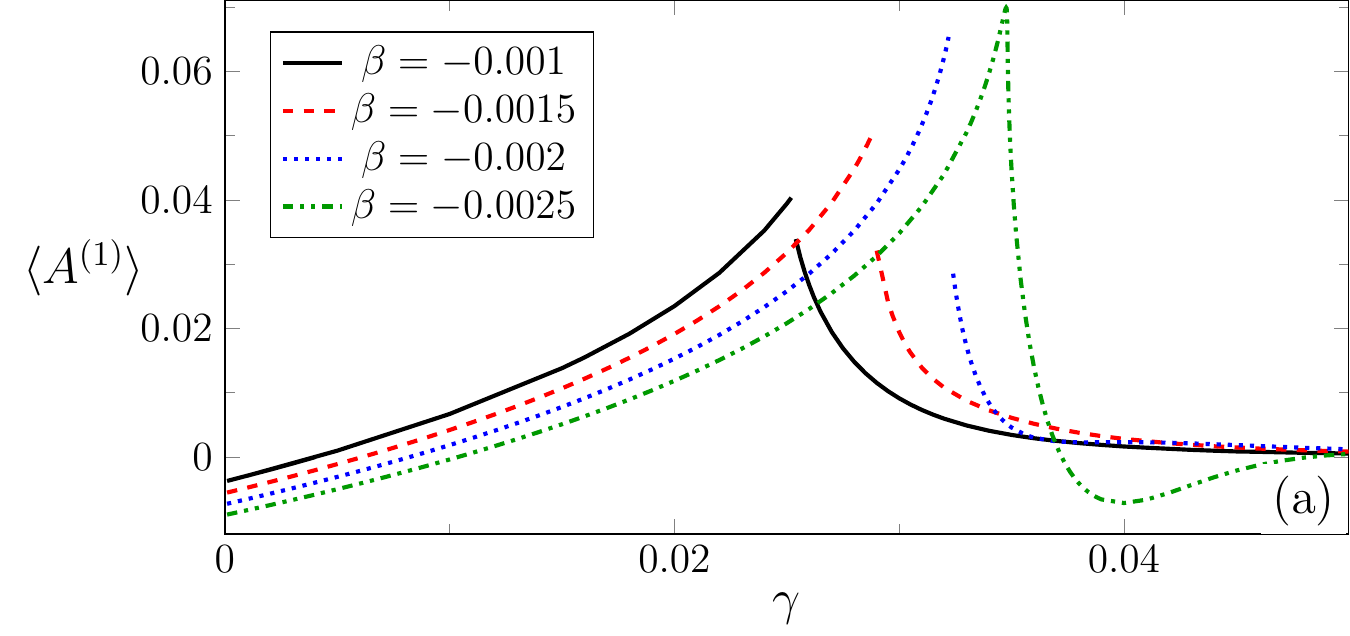} \\
    \hspace{-0.38cm}
    \includegraphics[scale=0.53]{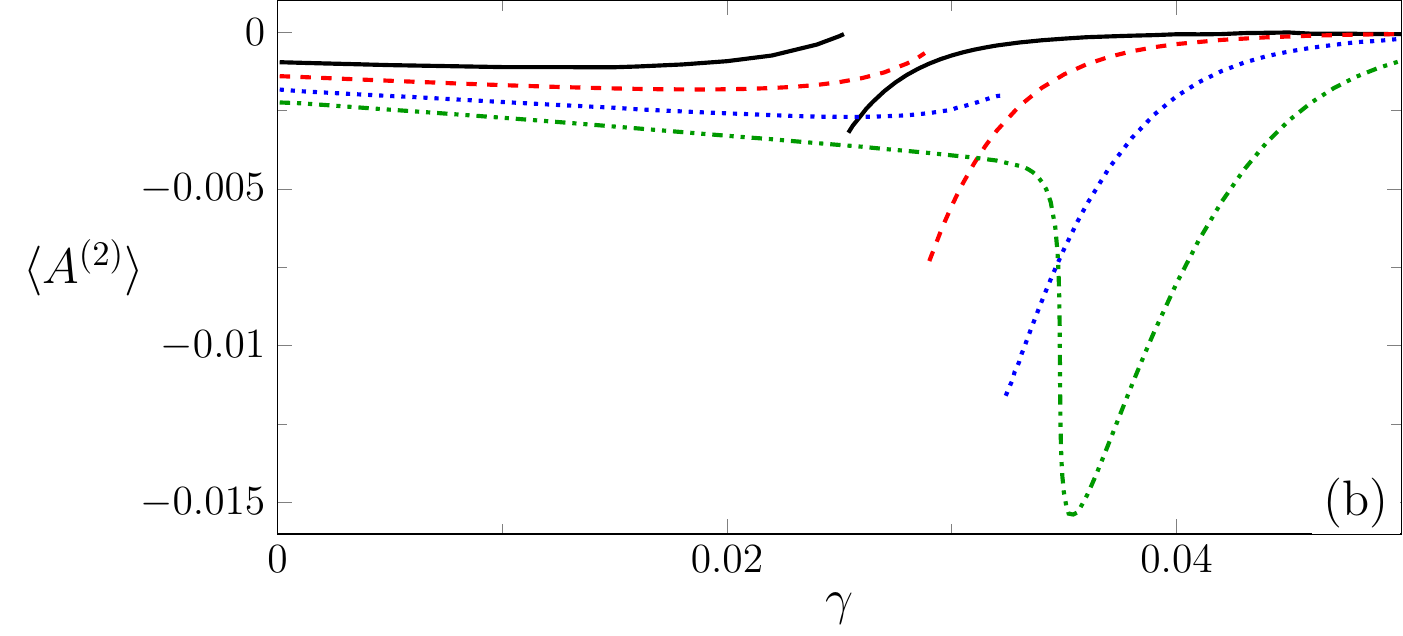}
    \caption{Degree assortativities $\langle A^{(1)} \rangle$ and $\langle A^{(2)} \rangle$ corresponding, respectively, to nearest neighbor nodes and
      next nearest neighbors nodes of an exponential random graph model with degree-degree correlations (see Eq. (\ref{hamil113})), $\beta<0$, and maximum degree $k_{\rm max} = 10$.}
\label{assortbubu}
\end{center}
\end{figure}

\begin{figure}[h]
  \begin{center}
    \includegraphics[scale=0.9]{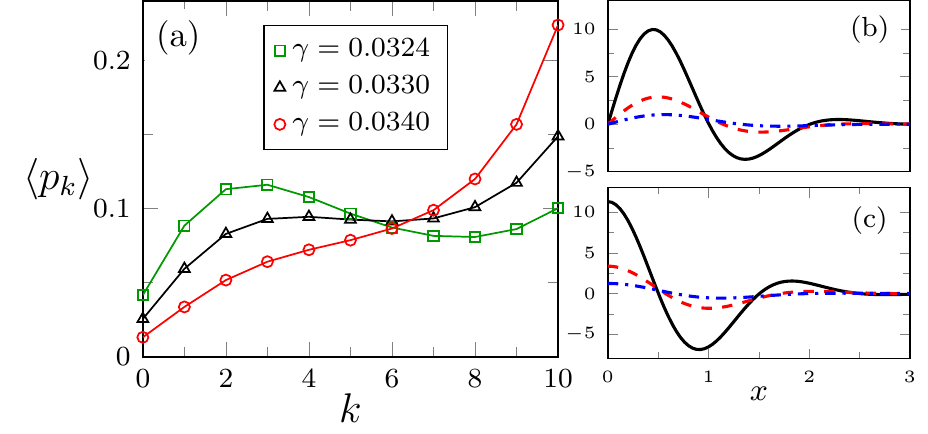}
    \caption{Degree distribution $\langle p_k \rangle$ and the order-parameter functions $\{ \rho_{k}^{*} (x) \}$ of an exponential random graph model with
      degree-degree correlations (see Eq. (\ref{hamil113})), $\beta = -0.002$, maximum
      degree $k_{\rm max} = 10$, and different values of $\gamma$ above the first-order transition.
      Figures (b) and (c) show, respectively, the functional behavior
      of ${\rm Im} \rho_{k}^{*} (x)$ and ${\rm Re} \rho_{k}^{*} (x)$ for $\gamma=0.0324$ and
      three values of $k$: $k=10$ (black solid lines), $k=9$ (red dashed lines), and  $k=8$ (blue dot-dashed lines). The other components $\rho_{0}^{*} (x), \dots, \rho_{7}^{*} (x)$
     are approximately zero.}
\label{assortatuta}
\end{center}
\end{figure}

In figure \ref{assortbubu} we present $\langle A^{(1)} \rangle$ and $\langle A^{(2)} \rangle$ as a function
of $\gamma$ for $\beta <0$. The ERG model undergoes a first-order transition at the critical point
$\gamma= \gamma_c(\beta)$.
For $\gamma < \gamma_c(\beta)$, the degree correlations do not considerably
affect $\langle p_k \rangle$, which is closer to a Poisson distribution, similar to figure \ref{diag}-(b). For $\gamma > \gamma_c(\beta)$, negative degree
correlations between next nearest neighbors have an important effect in the graph structure and $\langle p_k \rangle$ can exhibit a bimodal shape, as illustrated
in figure \ref{assortatuta}. For increasing $\gamma > \gamma_c(\beta)$, the weight $\langle p_{k_{\rm max}}  \rangle$  gradually increases
until we attain $\langle p_k \rangle = \delta_{k,k_{\rm max}}$ for $\gamma \rightarrow \infty$. Nevertheless, the graph
structure in the condensed phase for $\beta <0$ is qualitatively distinct from the regime $\beta >0$, which is also attested by the functional behavior of
the order-parameters, presented in figures \ref{assortatuta}-(b) and \ref{assortatuta}-(c) (as a comparison, see figure \ref{assortatuta1}-(b)). As shown by figure \ref{assortbubu}, the first-order transition
disappears for $\beta$ smaller than a certain threshold, which marks the terminating point of the first-order critical line in the plane $(\beta,\gamma)$.

\begin{figure}[h]
  \begin{center}
    \includegraphics[scale=0.6]{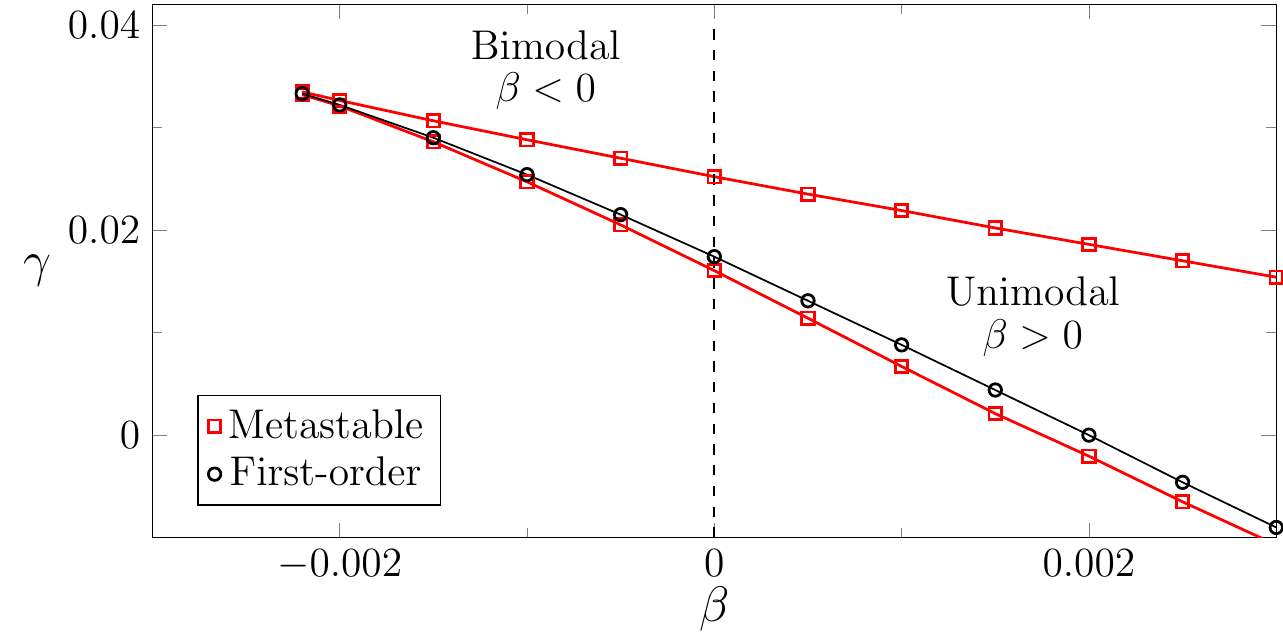}
    \caption{Phase diagram of an exponential random graph model with degree-degree correlations (see Eq.~(\ref{hamil113})) and maximum degree $k_{\rm max}=10$. The symbols
      are theoretical results obtained from the solutions of Eqs. (\ref{free}-\ref{sad2}), and the solid lines are just a guide to the eye. The phase diagram
      has a first-order critical line surrounded by a metastable region, where the free energy has two minima. Above the
      first-order critical line, the degree distribution is bimodal for $\beta <0$, and it has a single peak at the largest degree for $\beta >0$. The degree
      distribution approaches $\langle p_k \rangle = \delta_{k,k_{\rm max}}$ as $\gamma \rightarrow \infty$. 
}
\label{diagtuc}
\end{center}
\end{figure}

The competition between nearest neighbor and next nearest neighbor degree correlations is summarized in figure \ref{diagtuc}, which depicts the phase diagram of the
model defined by Eq. (\ref{hamil113}). The phase diagram exhibits a metastable region around a first-order critical
line $\gamma_c(\beta)$ that terminates at a negative value of $\beta$. For $\gamma < \gamma_c(\beta)$, $\langle p_k \rangle$ is closer
to a Poisson distribution. For $\gamma > \gamma_c(\beta)$ and $\beta > 0$,  we have $\langle A^{(2)} \rangle > 0$
and the degree distribution has a single peak at $k=k_{\rm max}$. For $\gamma > \gamma_c(\beta)$ and $\beta < 0$, we have $\langle A^{(2)} \rangle < 0$
and $\langle p_k \rangle$ can exhibit two peaks, one of them located at $k=k_{\rm max}$, and an additional peak at a smaller degree.
%


\section{Final remarks}

In this paper we have solved the two-star model with degree-degree correlations in
the sparse regime. The model allows to generate random graphs with prescribed degree correlations
between adjacent nodes and between nodes at the end-points of two-stars. By introducing an upper cutoff
in the degree sequence, we have exactly calculated the network free energy, from which we
derived complete phase diagrams and characterized the graph structure in the different phases.

In terms of the degree distribution $\langle p_k \rangle$, the phase diagram of the model is characterized
by three distinct regions.
There is a phase where $\langle p_k \rangle$ is approximately given
by a Poisson distribution, reminiscent from the structure of Erd\"os-R\'enyi random graphs \cite{Newman10}. The
phase diagram also exhibits a condensed phase, where the shape of $\langle p_k \rangle$ strongly depends on the 
degree correlations. If the degree assortativities are non-negative inside the condensed phase, then $\langle p_k \rangle$ has a single peak
at the maximum degree and the graph is approximately regular. If the degree assortativity of next
nearest neighbors is negative inside the condensed phase, then $\langle p_k \rangle$ is given by a bimodal distribution, with one
maximum at the largest degree and an additional maximum at a smaller degree. While the Poisson and standard condensed phases appear even in the
absence of degree-degree correlations, the existence of a bimodal degree distribution is a genuine effect of negative degree correlations between next
nearest neighbor nodes. This result reveals the importance of long-range degree correlations, beyond nearest neighbor nodes, in shaping the
network structure.

We have shown that the model undergoes a first-order transition between the Poisson phase
and the condensed phase.
The first-order critical line is surrounded by a metastable region, where the free energy has two minima, each one corresponding
to a phase. For combinations of model parameters inside the metastable region, algorithms to sample finite graphs from this ERG model may get stuck
in a local minimum of the free energy \cite{CoolenBook}.
In addition, the jump of the structural observables across the first-order critical line prevents us from generating graph samples
with structural parameters in a certain range. Taken together, these features represent serious limitations of the present ERG model as an effective tool to model real-world networks.
The analytic solution of the model for $N \rightarrow \infty$ and the construction of its phase diagrams have practical
relevance, as these results allow to estimate the metastable regions in the parameter space of finite graphs.

The present paper constitutes a first step towards controlling the generation of ERGs with correlated degrees.
Overall, the results for the
assortativities, the degree distribution, and the phase diagrams allow to identify 
the regime of parameters where the model can be useful to reproduce certain properties of empirical networks.
A drawback of the present model is that the degree distribution in each phase does not bear any resemblance to the broad degree distributions found in real-world networks. With the purpose
of improving the model, it would be interesting to solve it with a hard constraint in the degree sequence or with a prescribed
degree distribution \cite{Cimini2019}. 
This work also opens the perspective to explore systematically the role of short-range and long-range degree correlations in
dynamical processes occurring on tree-like networks, since in this case the equations for the dynamics are typically determined
only by the degree distribution \cite{Edgar2020}. Finally, we point out that the free energy of ERG models can be mapped in the cumulant
generating function of certain structural observables of Erd\"os-R\'enyi random graphs \cite{Metz2016,Metz2019,giardina2021} . Therefore, the results of the present paper can be
readily applied to study analytically the large deviations of higher-order topological properties of Erd\"os-R\'enyi random graphs
in the limit $N \rightarrow \infty$ \cite{Chen2019}. 

\section*{acknowledgements}

M.B. acknowledges a fellowship from CAPES/Brazil. F.L.M. and I.P.C. gratefully acknowledge London Mathematical Laboratory
for financial support. F.L.M. also acknowledges a fellowship from CNPq/Brazil


\appendix

\section{Symmetry properties of the order-parameter functions for $\beta < 0$}

In this appendix we obtain the symmetry properties of the order-parameters $\rho_0^{*} (x),\dots, \rho_{k_{\rm max}}^{*} (x)$ under
the transformation $x \rightarrow -x$. These properties allow to simplify the saddle-point Eqs. (\ref{sad1}) and (\ref{sad2}) and the computation
of the structural observables introduced in section \ref{observables}.

For $\beta < 0$,  $\{  \rho_k^{*} (x),\hat{\rho}_k^{*} (x) \}$ are complex-valued functions of $x \in \mathbb{R}$. This can be
seen from Eq. (\ref{sad1}), which can be written for $\beta <0$ as
\begin{equation}
  \hat{\rho}_{k}^{*}(x)  =  \sum_{l=0}^{k_{\rm max}} U_{kl}  e^{i \sqrt{|\beta|} x l }
  \int_{-\infty}^{\infty} d x^{\prime} \rho_{l}^{*}(x^{\prime})  e^{i \sqrt{|\beta|} x^{\prime} k } , \label{sjos}
\end{equation}  
where
\begin{equation}
U_{kl} = e^{\gamma D(k,l) + \frac{|\beta|}{2} \left( k^2 + l^2  \right)  }.
\end{equation}  
If we take the complex-conjugate $\overline{\left( \dots  \right)}$ of Eq. (\ref{sad2}) and make the
transformation $x \rightarrow - x$, then we get
\begin{equation}
 \overline{\rho_{k}^{*}}(-x) =   \frac{1}{\overline{\mathcal{T}}} \frac{k}{k!}  \left[ \, \overline{\hat{\rho}_{k}^{*}}(-x) \right]^{k-1}   e^{ -\frac{1}{2} x^2 +  \sum_{r=1}^Q \alpha_r F_r(k) }      , \label{kopa}
\end{equation}  
with $\overline{\mathcal{T}}$
\begin{equation}
  \overline{\mathcal{T}} =  \sum_{k=0}^{k_{\rm max}}  \frac{1}{k!}     \int_{-\infty}^{\infty}  d x   \left[ \, \overline{\hat{\rho}_{k}^{*}}(-x) \right]^k   e^{ -\frac{1}{2} x^2 +   \sum_{r=1}^Q \alpha_r F_r(k) }.
  \label{normal}
\end{equation}  
By taking the complex-conjugate of Eq. (\ref{sjos}), the function $\overline{\hat{\rho}_{k}^{*}}(-x)$ fulfills
\begin{equation}
  \overline{\hat{\rho}_{k}^{*}}(-x)  =  \sum_{l=0}^{k_{\rm max}} U_{kl}  e^{i \sqrt{|\beta|} x l }
  \int_{-\infty}^{\infty} d x^{\prime} \, \overline{\rho_{l}^{*}}(-x^{\prime})  e^{i \sqrt{|\beta|} x^{\prime} k } . \label{sjos1}
\end{equation}  
Since Eqs. (\ref{sad1}) and (\ref{sad2}) for $\{  \rho_k^{*} (x),\hat{\rho}_k^{*} (x) \}$ are the same
as Eqs. (\ref{kopa}) and (\ref{sjos1}) for $\{ \overline{\rho_k^{*}}(-x),\overline{\hat{\rho}_k^{*}}(-x) \}$, we conclude
that
\begin{equation}
  \rho_k^{*} (x) = \overline{\rho_k^{*}}(-x)
  \label{pari1}
\end{equation}  
and
\begin{equation}
  \hat{\rho}_k^{*} (x) = \overline{\hat{\rho}_k^{*}}(-x)
  \label{pari2}
\end{equation}  
for arbitrary $x$. This implies that $\{  {\rm Re} \rho_k^{*} (x),{\rm Re} \hat{\rho}_k^{*} (x) \}$ and
$\{  {\rm Im} \rho_k^{*} (x),{\rm Im} \hat{\rho}_k^{*} (x) \}$ are, respectively, even and odd functions
of $x$.

Let us use the above symmetry properties to simplify the order-parameter equations. By setting $\hat{\rho}_k^{*} (x)$ in polar
form
\begin{equation}
\hat{\rho}_k^{*} (x) = r_k(x) e^{i \varphi_k (x)}, 
\end{equation}  
with  $\varphi_k (x) \in (-\pi,\pi]$, and noting that
\begin{equation}
r_k(-x) = r_k(x)
\end{equation}  
and
\begin{equation}
\varphi_k (-x) =  -\varphi_k (x),
\end{equation}  
we readily obtain ${\rm Im}\mathcal{T} = 0$ from Eq. (\ref{normal}).
Hence Eq. (\ref{sad2}) can be written as
\begin{align}
  {\rm Re} \rho_{k}^{*}(x) &=
  \frac{1}{ {\rm Re} \mathcal{T} } \frac{k}{k!}  \left[ r_k(x) \right]^{k-1}  \cos{\left[ (k-1) \varphi_k(x) \right] }  \nonumber \\
  &\times e^{ -\frac{1}{2} x^2 +  \sum_{r=1}^Q \alpha_r F_r(k) }      , \label{popa1} \\
   {\rm Im} \rho_{k}^{*}(x) &=
  \frac{1}{ {\rm Re} \mathcal{T} } \frac{k}{k!}  \left[ r_k(x) \right]^{k-1}  \sin{\left[ (k-1) \varphi_k(x) \right] }  \nonumber \\
  &\times e^{ -\frac{1}{2} x^2 +  \sum_{r=1}^Q \alpha_r F_r(k) }      . \label{popa2}
\end{align}  
The above equations are coupled to Eq. (\ref{sad1}), which can be simplified for $\beta < 0$ using Eqs. (\ref{pari1}) and (\ref{pari2}) 
\begin{align}
&{\rm Re} \hat{\rho}_{k}^{*}(x) = 2 \sum_{l=0}^{k_{\rm max}} U_{kl} \cos{\left( \sqrt{|\beta|} x l \right)}  \int_{0}^\infty d x^{\prime} Y_{kl}(x^{\prime}),  \label{popa3}  \\
  &{\rm Im} \hat{\rho}_{k}^{*}(x) = 2 \sum_{l=0}^{k_{\rm max}} U_{kl} \sin{\left( \sqrt{|\beta|} x l \right)}  \int_{0}^\infty d x^{\prime}  Y_{kl}(x^{\prime}), \label{popa4} 
\end{align}
where
\begin{eqnarray}
Y_{kl}(x) &=&  {\rm Re} \rho_{l}^{*}(x^{\prime})    \cos{\left( \sqrt{|\beta|} x^{\prime} k \right)} \nonumber \\
    &-&  {\rm Im} \rho_{l}^{*}(x^{\prime}) \sin{\left( \sqrt{|\beta|} x^{\prime} k \right)}. 
\end{eqnarray}  
The fixed-point functions $\{ \rho_{k}^{*}(x) \}$ that solve Eqs. (\ref{popa1}-\ref{popa4}) determine the free energy $f$ and all the structural
parameters for $\beta < 0$. Using the symmetry properties of Eqs. (\ref{pari1}) and (\ref{pari2}), it is straightforward to verify
from Eq. (\ref{jojo}) that $f \in \mathbb{R}$ for $\beta < 0$.

\bibliography{biblio.bib}

\end{document}